\pdfoutput=1

\documentclass[nohyper]{JINST}

\graphicspath{{figures/}}

\title{The H1 Forward Track Detector at HERA II}

\author{P.J. Laycock$^a$\thanks{Corresponding
author.}~, R.C.W. Henderson$^b$, S.J. Maxfield$^a$, J.V. Morris$^c$, G.D. Patel$^a$ and D.P.C. Sankey$^c$\\
\llap{$^a$}University of Liverpool,\\
 Liverpool L69 3BZ, UK\\
\llap{$^b$}Lancaster University,\\
  Lancaster LA1 4YB, UK\\
\llap{$^c$}Rutherford Appleton Laboratory,\\
  Didcot, Oxon., OX11 0QX, UK\\
  Corresponding Author E-mail: \email{laycock@hep.ph.liv.ac.uk}}

\abstract{
In order to maintain efficient tracking in the forward region of H1
after the luminosity upgrade of the HERA machine, the H1 Forward Track
Detector was also upgraded. While much of the original software and
techniques used for the HERA I phase could be reused, the software for
pattern recognition was completely rewritten. This, along with several
other improvements in hit finding and high-level track reconstruction, are
described in detail together with a summary of the performance of the
detector.}


\begin{document}

\section{Introduction}

The original HERA I incarnation of the H1 Forward Track Detector (FTD)
is described in detail in \cite{Burke:1995de} and much of the
information found there remains relevant for the HERA II FTD. However,
the luminosity upgrade of the HERA machine in 2000-2002 required a
significant upgrade of the FTD in order to accommodate the higher
particle multiplicities and maintain high track-finding efficiency and
purity; it is these various improvements which are described here. A
brief description of the upgraded FTD hardware, which took data
between 2002-2007, is given in the next section. Section 3 addresses
the hit-finding modifications which were made in order to accommodate
the higher currents in the FTD and mitigate the effects ageing in the
chambers. In Section 4, a complete description of the new pattern
recognition algorithms is given, while section 5 details the higher
level pattern recognition modification using the Kalman
Filter. Section 6 describes the combination of the FTD information
together with information from the central trackers.  Finally, Section
7 describes the performance of the FTD.

\section{Overview of the HERA II FTD}

\begin{figure}[h]
\begin{center}
\includegraphics[width=0.7\columnwidth]{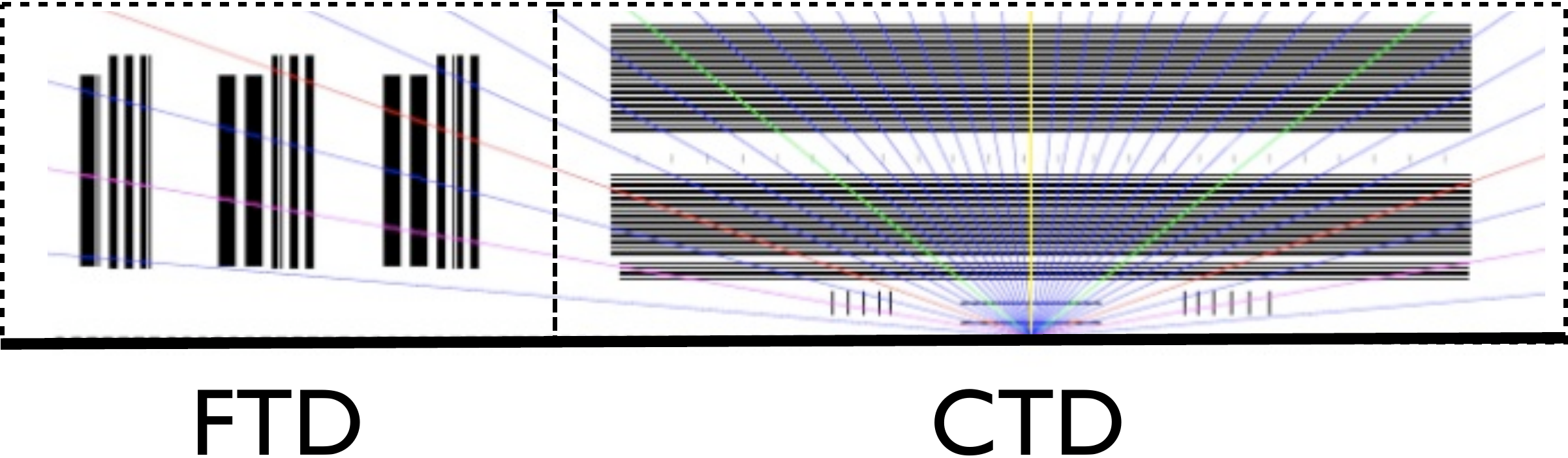}
\caption{
A schematic showing the (top half of the) FTD together with the
central H1 tracking detectors, labelled CTD.  The silicon trackers are
located closest to the beam pipe (indicated by the thick black line),
the nominal interaction point is at the focus of the lines showing $5$
degree intervals in the polar angle (figure courtesy of Daniel
Pitzl).}

\label{Fig:H1tracking}
\end{center}
\end{figure}

A description of the H1 detector can be found elsewhere \cite{H1Det}.
Figure $\ref{Fig:H1tracking}$ shows the FTD together with the tracking
detectors in the central region of the H1 detector, referred to here
as the H1 Central Track Detector or CTD.  The CTD consists of several
detector technologies \cite{H1Det}, including forward, central and
backward silicon detectors which can be seen located close to the beam
pipe \cite{H1Silicon}.

The original FTD \cite{Burke:1995de} comprised various detector
technologies: radial and planar draft chambers, multi-wire
proportional chambers and transition radiators. The upgraded FTD
reused the planar draft chambers, referred to hereafter as P-modules,
and replaced all other detectors with new planar drift chambers or
Q-modules. Each module has sense wires arranged parallel to one
another in planes perpendicular to the $z$-axis\footnote{The origin of
the H1 coordinate system is the nominal $ep$ interaction point with
the direction of the proton beam defining the positive $z$-axis
(forward direction). The polar angle~($\theta$) is defined with
respect to this axis and the pseudorapidity is defined as
$\eta=-\ln{\tan(\theta/2)}$. The azimuthal angle $\phi$ defines the
particle direction in the transverse plane.}  at positions between
$130$ and $240$ cm from the nominal interaction point.

\begin{figure}[h]
\begin{center}
\includegraphics[width=1.0\columnwidth]{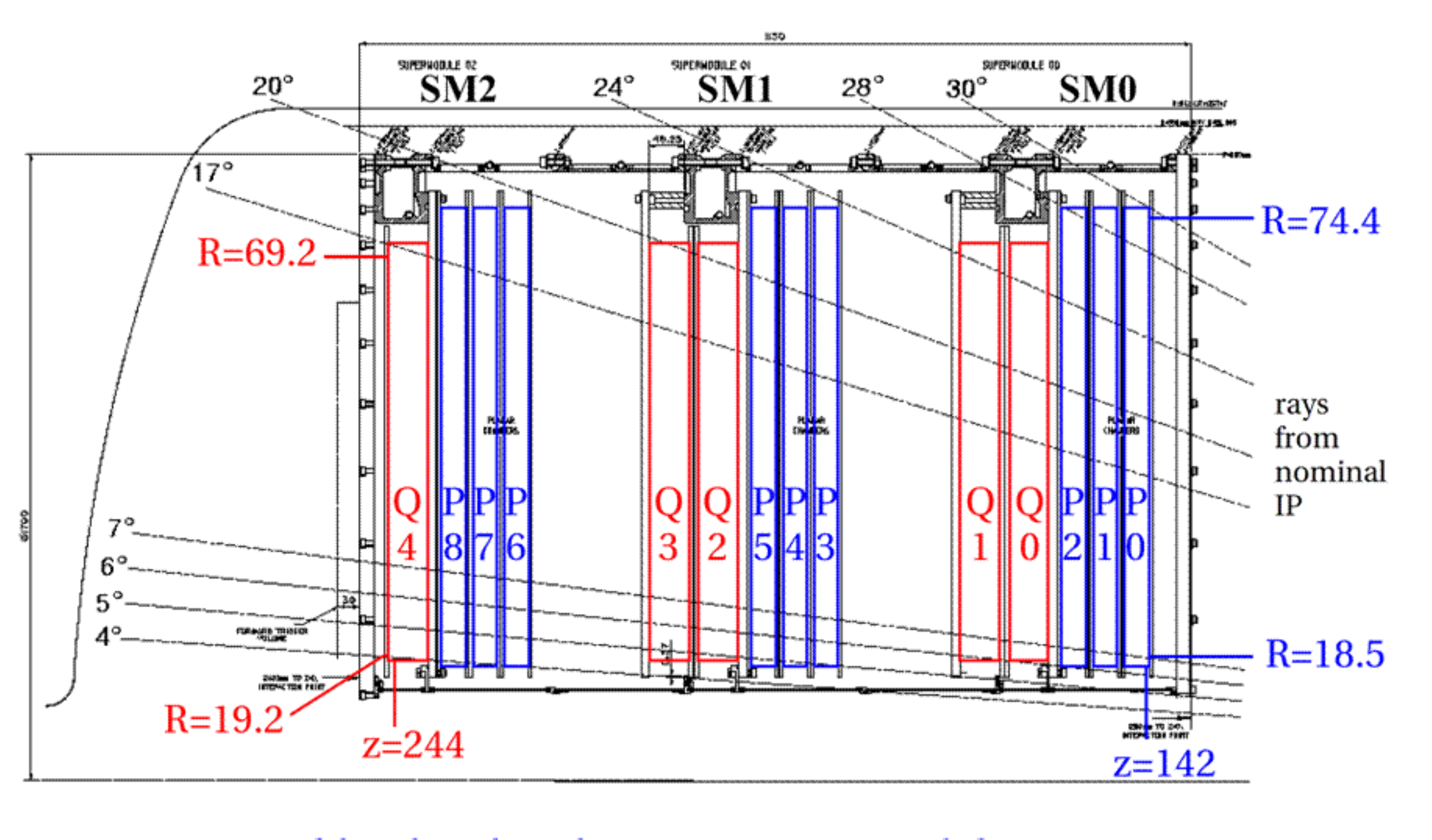}
\caption{A schematic cross-section of (the top half of) the FTD
  showing the arrangement of modules(P, Q) within the three
  supermodules (SM0, SM1 and SM2). The interaction point is to the
  right. }
\label{Fig:FTDside}
\end{center}
\end{figure}

\begin{figure}[h]
\begin{center}
\includegraphics[width=0.9\columnwidth]{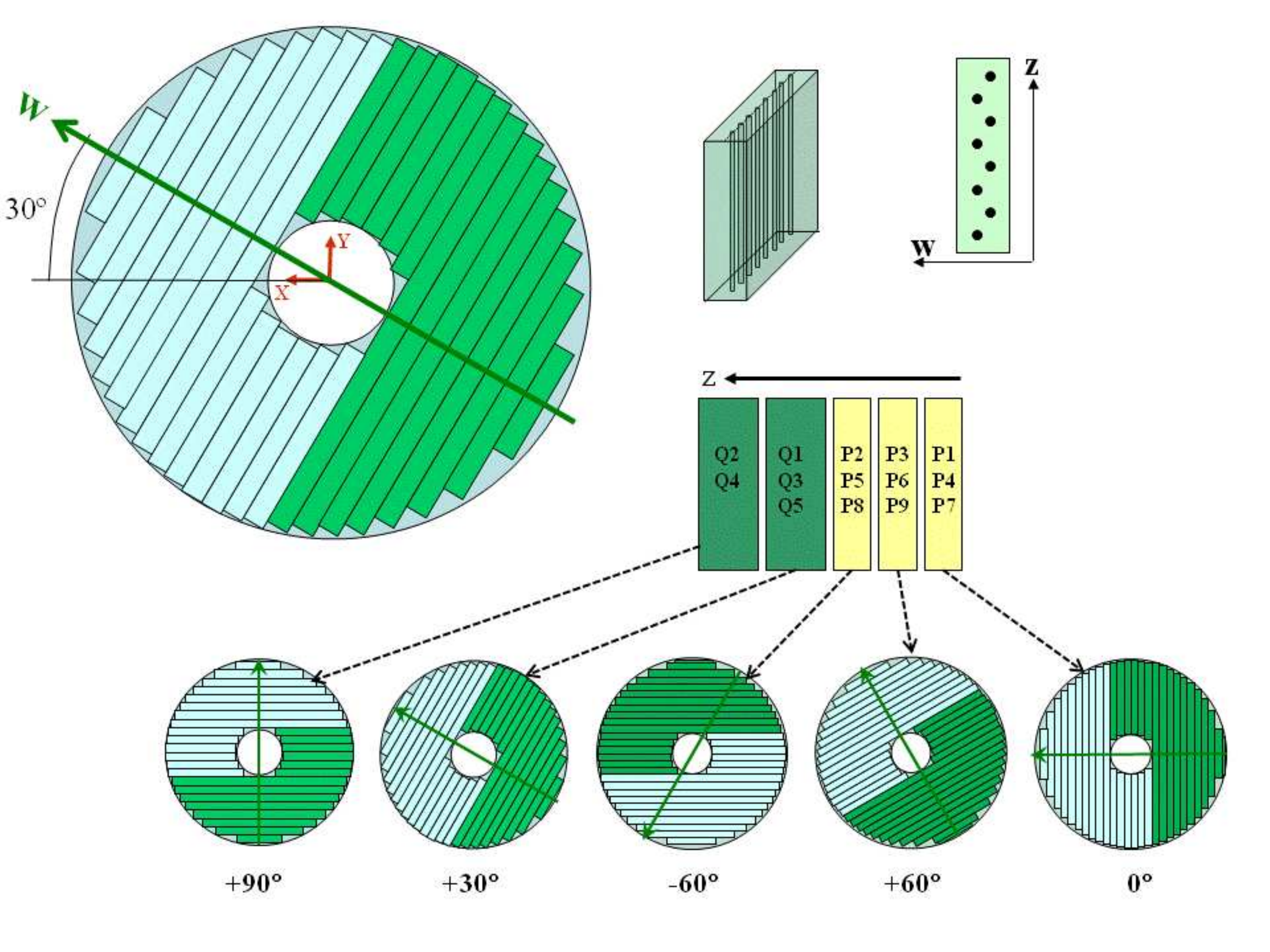}
\caption{A schematic view of one supermodule of the FTD showing the
  orientation of wires in successive modules.}
\label{Fig:FTDorientations}
\end{center}
\end{figure}

The modules were grouped into three "supermodules", as shown in figure
$\ref{Fig:FTDside}$. The first two supermodules contained three
P-modules and two Q-modules and the third, furthest from the
interaction point, had three P-modules and one Q-module.  Within a
supermodule, the modules were rotated in $\phi$ with respect to one
another so that the combined drift coordinate measurements could be
used to locate tracks in space. The wires in the first module were
parallel to the $y$-axis. Subsequent modules were rotated by
$+60^{\circ}$, $-60^{\circ}$, $+30^{\circ}$ and $+90^{\circ}$ with
respect to this (see figure $\ref{Fig:FTDorientations}$). In the
following, the drift coordinate $w$ is the coordinate perpendicular to
the sense wires. Thus the $w$-axis coincides with the $x$-axis in the
first module and is rotated with respect to the $x$-axis by the angles
given above in the others. The three supermodules were enclosed in a
gas tank through which circulated a $Ar/C3H8$ gas mixture and which
also served as a mechanical support.

The wires in each plane spanned disc-shaped regions of approximately
$80$ cm radius. A concentric $15$ cm radius hole in the disc
accommodated the beam pipe. Wires that would otherwise cross this hole
were split into two separate parts. Modules contained either four wire
planes (P-modules) or eight wire planes (Q-modules) separated in $z$
by $0.6$~cm. In P modules each wire plane had $32$ wires spaced
$5.7$~cm apart across the plane.  Q modules had $28$ wires per wire
plane, $6.2$~cm apart. The drift field was maintained by a set of
field wires and cathode planes surrounding the sets of four
(P-modules) or eight (Q-modules) sense wires, as shown in figure
$\ref{Fig:FTDorientations}$. Note that the wires were given a $\pm
300\mu$m offset ("stagger") from their nominal locations in order to
help resolve "left-right" ambiguities. The wires were read out at one
end only, providing a measurement of the drift coordinate
perpendicular to them.

It should be noted that a significant fraction of the new Q-modules
had severe hardware problems, with $25\%$ of the Q-modules in the
first supermodule and $50\%$ of the only Q-module in the third
supermodule being completely inoperable.  These and other smaller
hardware failures were masked out of the pattern recognition both for
real data and simulation.  Thanks to the redundancy built into the
design of the FTD, the overall performance of the detector was robust
despite the impact on pattern recognition at the supermodule-level.

\section{Readout and Hit Finding}

\subsection{Introduction}

As with the original HERA I FTD, the sense wires in the HERA II FTD
were read out using $104$~MHz $8$ bit non-linear Flash
Analogue-to-Digital Converters (FADCs) giving a history of the chamber
pulses in time-slices of $9.6$~ns. The wires were instrumented at one
end only. Hits, which correspond to the ionisation left behind by a
charged particle passing through the detector, were found using a QT
algorithm which analysed the charge and time information of the
chamber pulses.  Given the similarity between the two
types of modules, the same QT algorithm was used in both to detect
hits.  The original version of the QT algorithm was based on that
used for the old P-modules, described in \cite{Burke:1995de}, with the
further modification of different thresholds for the "outer" wires,
the wires with only one adjacent sense wire, in each set of
chambers. These thresholds were implemented as an additional
multiplicative factor in the relinearisation lookup table, with a
nominal value of unity for the inner wires (in the Q-modules, the middle six wires)
and a different factor for the outer
wires. The thresholds for the outer wires were downscaled by a factor of $1.9$ and $3.0$
in the P and Q-modules, respectively, the difference arising from the
different physical geometry and therefore electrostatics of the new
modules. 

\begin{figure}[h]
\begin{center}
\includegraphics[width=0.49\columnwidth]{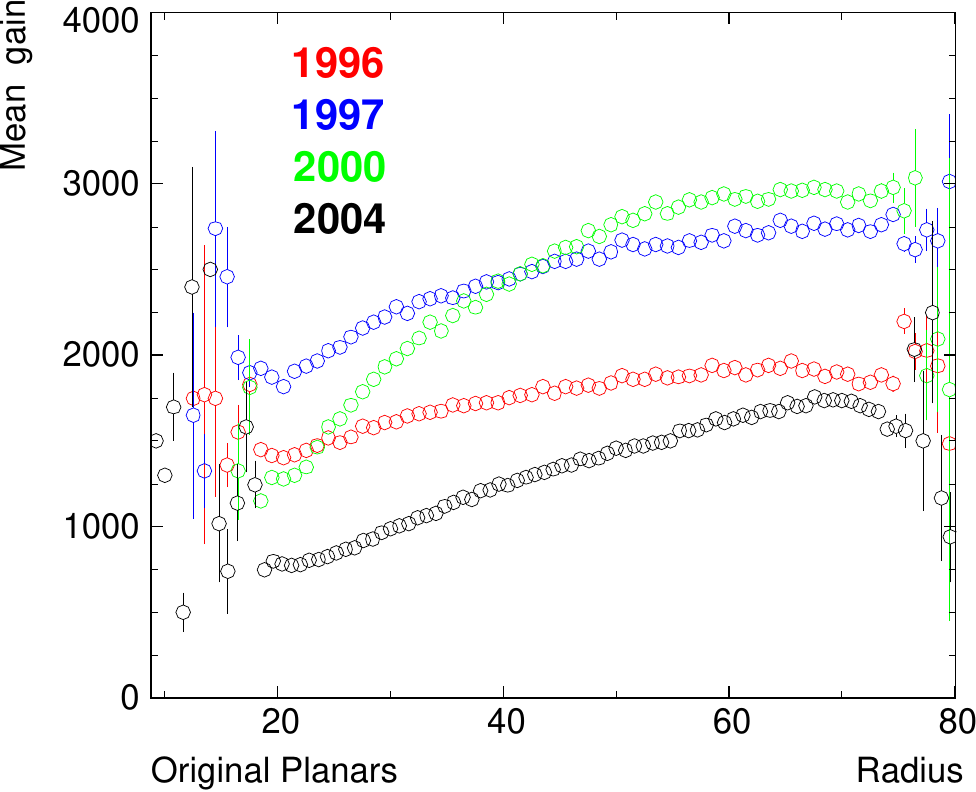}
\includegraphics[width=0.47\columnwidth]{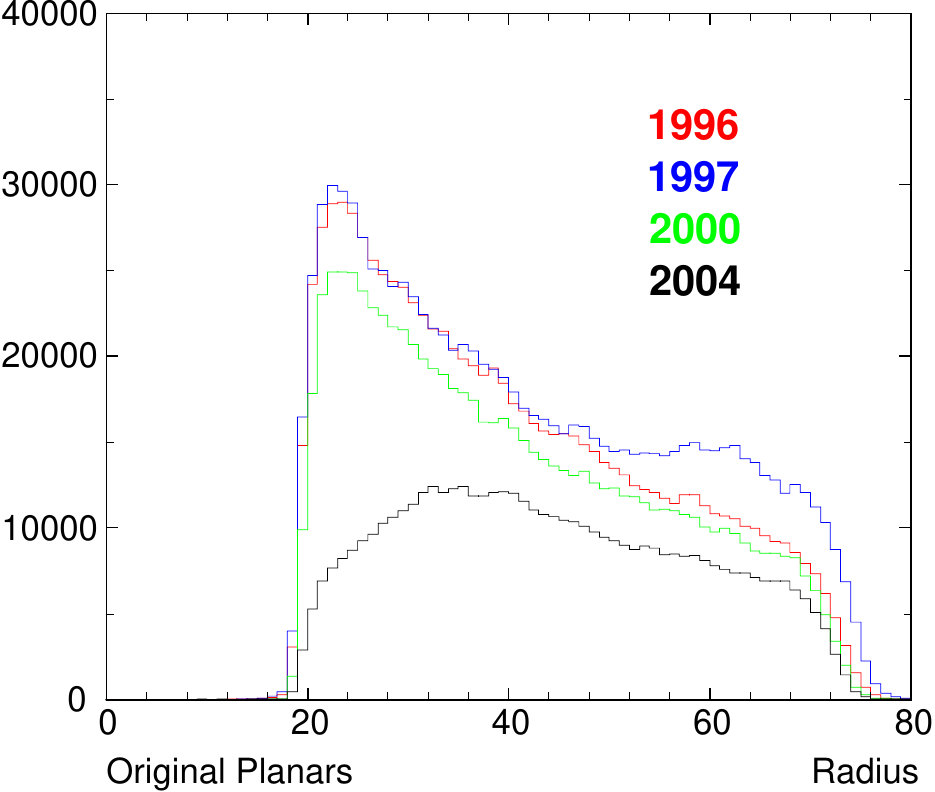}
\caption{Mean pulse height as a function of radius for the P-modules for representative runs in
$1996$, $1997$, $2000$ and $2004$ (left) and the number of hits as a function of radius
for the same runs (right).}
\label{Fig:QT}
\end{center}
\end{figure}

During commissioning of the FTD with collision data, the observed
single-wire efficiency of the P-modules was found to be markedly lower
than expected. Extensive investigations ruled out problems with the
gas mixture\footnote{In addition to the $Ar/C3H8$ gas mixture, ethanol
was a standard component of the gas mixture at the level of $\sim1\%$.
Increasing the ethanol content had no effect on the gain, although the
tests revealed that the operational stability of the chambers was indeed far better with
the inclusion of ethanol.} or oxygen contamination\footnote{The drift time
distribution was approximately flat, with known features caused by the electrostatics of
the cells.  There was no evidence for an inefficiency increasing with drift distance
as would be expected if there was oxygen contamination.  A dedicated
monitor also confirmed that any possible oxygen contamination was negligible.} 
and finally revealed the first
signs of ageing. Following a reanalysis of the HERA I data, it became
apparent that ageing in the P-modules had already been occurring
during HERA I running, but the effects of this had been masked by
increases in the gain in the chambers such that the resulting
inefficiencies were minimal. Upon discovering ageing,
the gain (high voltage) was lowered in order to preserve the chambers. However, the
result was that the P-modules became markedly inefficient at small
radius. The mean pulse height in the P-modules as a
function of radius is shown in figure $\ref{Fig:QT}$ for collision data-taking runs in
$1996$, $1997$, $2000$ and also in $2004$ (after the reduction in gas gain). 
The number of hits as a function of radius is also shown in
figure $\ref{Fig:QT}$
for the same runs, where the loss of hits at small radius is
immediately apparent for the $2004$ data. From the former, it is apparent
that the onset of degradation was as early as $1997$.

A new QT algorithm
was then developed over the next year and implemented in $2006$. This
algorithm was a development of that described for the Radial chambers
in \cite{Burke:1995de}, adapted for single-ended readout. In
particular the time determination, charge determination (save for the
integration interval being $8$ time-slices) and overlapping hit analysis
were as described previously. The major changes were in the hit
detection and noise suppression, in order to enhance detection of
small hits but reduce the sensitivity to noise, including the $10$~MHz
pickup which was an issue in some cells (in the HERA I Radial chambers
this was circumvented in the hit detection by summing the signal from
each end of the wire prior to the hit search). 

\begin{figure}[h]
\begin{center}
\includegraphics[width=0.6\columnwidth]{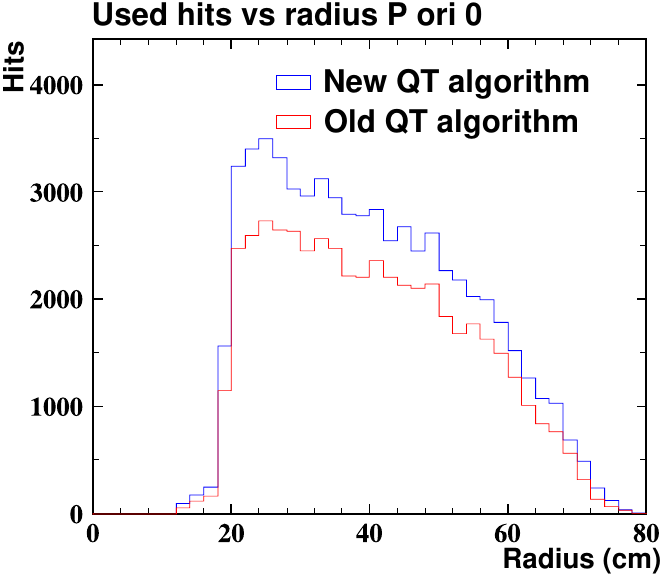}
\caption{The distribution of used hits as a function of radius for the
P-modules nearest the interaction point in run $435774$.}
\label{Fig:QThitscomparison}
\end{center}
\end{figure}

The success of these changes can be judged from the new algorithm
producing $5\%$ fewer hits in the P-modules and $20\%$ fewer in the
Q-modules.  These hits form more pattern-recognised ``clusters'',
which are groups of contiguous hits in a module (see section
$\ref{subsec:clusters}$).  There are $20\%$ more hits used in P-module
clusters and $5\%$ more in the Q-modules when compared with the
previous algorithm, as illustrated in figure
$\ref{Fig:QThitscomparison}$ which shows the distribution of used hits
as a function of radius for the P-modules nearest the interaction
point in run $435774$. The changes are detailed in the following.

\subsection{Modifications to the algorithm}

First, crosstalk compensation was applied using a combination of the
signal and difference of samples signal ($DOS$, defined as
$DOS(n)=FADC(n)-FADC(n-1)$) of adjacent wires. Then a multistep adaptive
hit search was performed, processing each wire consecutively within a
block of four channels or wires (corresponding to a complete cell in
the P-modules and a half-cell in the Q-modules). First, the
data on the wire are scanned for potential hits consisting of a region
of one or more time slices of increasing charge, corresponding to the
leading edge of a potential pulse. If the total increase in charge is
above a threshold then the size, location and number of such regions
is recorded. At this stage, if there are no such regions in the data
from a single channel, then a single small hit is accepted if its
increase in charge is greater than half the current threshold; the
threshold is then modified accordingly. Next, the hit threshold is
scaled up as a function of the number of potential hits found on the
wire, such that the hit search is automatically hardened in the
presence of multiple hits (or in particular, oscillating noise). The
potential hit candidates are compared to this modified threshold (and
in the case of more than one candidate, the additional requirement
that there are at least two time slices on the leading edge of the
pulse).

\begin{figure}
\begin{center}
\includegraphics[width=0.8\columnwidth]{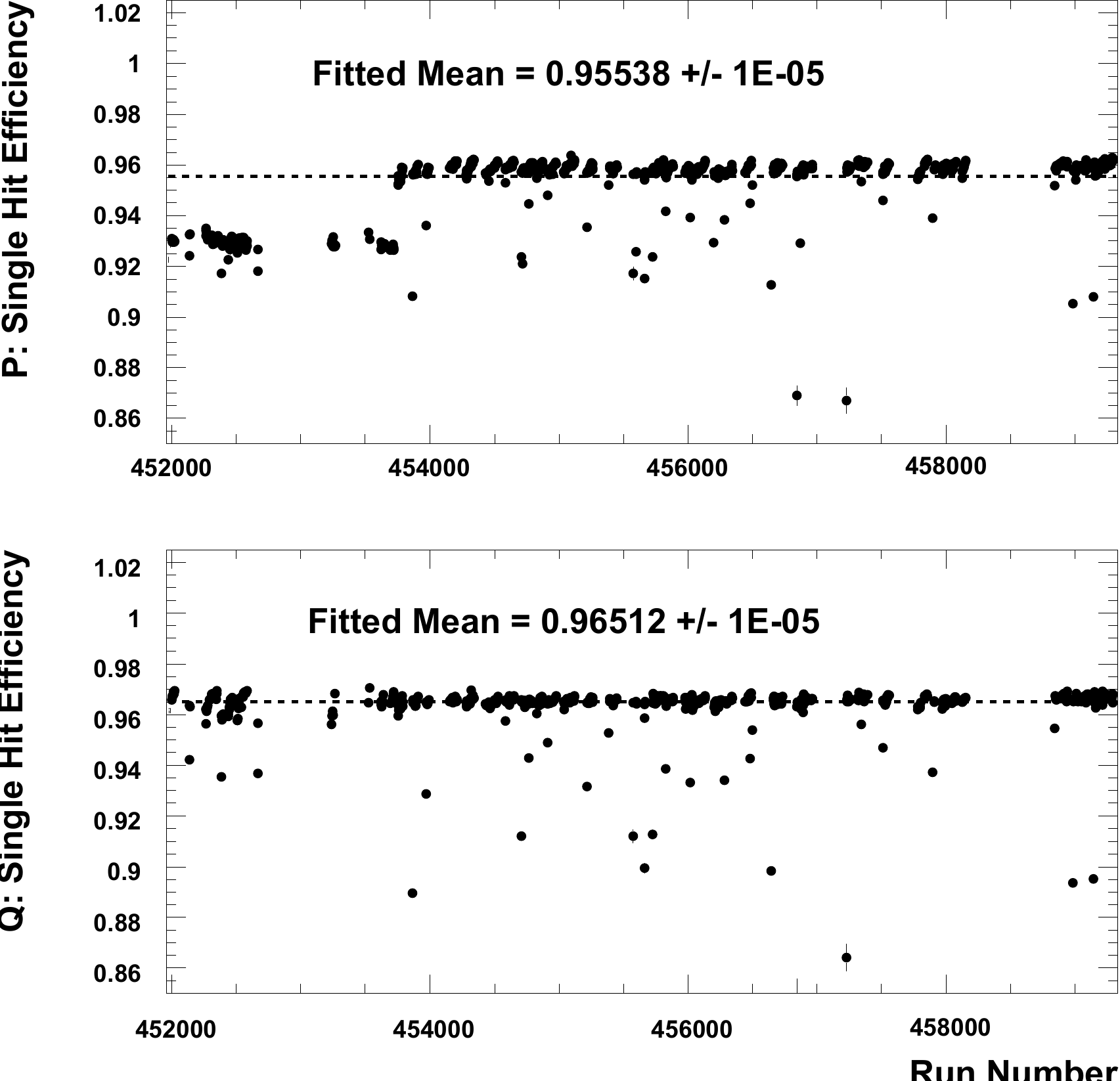}
\caption{The monitored single hit efficiency of the old P-modules, the change to the new
QT algorithm can be clearly seen.}
\label{Fig:QTefficiency}
\end{center}
\end{figure}

The charge is then determined for hits passing this modified
threshold. If this charge is greater than twice the threshold the hit
is accepted (again effecting protection against oscillating
noise). Finally there is a requirement that in each block of four
wires there is at least one hit above the original full threshold,
failing which all hits for the wires are suppressed, as a final
protection against noise. This multistep search allows a significantly
lower hit threshold to cope with inefficient wires whilst maintaining
robust protection against noise. For the P-modules, the hit
threshold used was half that used in the original algorithm, while for
the Q-modules it was $7/8$. The smallest hits that could be detected
were therefore a quarter of the size of the minimum with the original
algorithm. These changes restored the monitored single hit efficiency
of the old P-modules to around $96\%$, matching the Q-modules, as
illustrated by figure $\ref{Fig:QTefficiency}$ taken from the on-line
monitoring at the time the new algorithms went live.  The successful
conclusion of these changes allowed stable operation of the FTD for the
remainder of the HERA II running, with an uptime of $\sim98\%$.

\section{H1 FTD Pattern Recognition}

\subsection{Introduction}

The pattern recognition for the H1 FTD is achieved in distinct
stages. Firstly, the hits found by the QT algorithm in Section 3 are
combined into clusters of hits in individual modules. These clusters
are then combined at the level of supermodules to make track
segments. Finally, those segments are linked together and fit using a
modified version of the Kalman filter described in
\cite{Burke:1995de}.

\subsection{Clusters}
\label{subsec:clusters}
\subsubsection{Finding candidate clusters}

\begin{figure}
\begin{center}
\includegraphics[width=0.52\columnwidth]{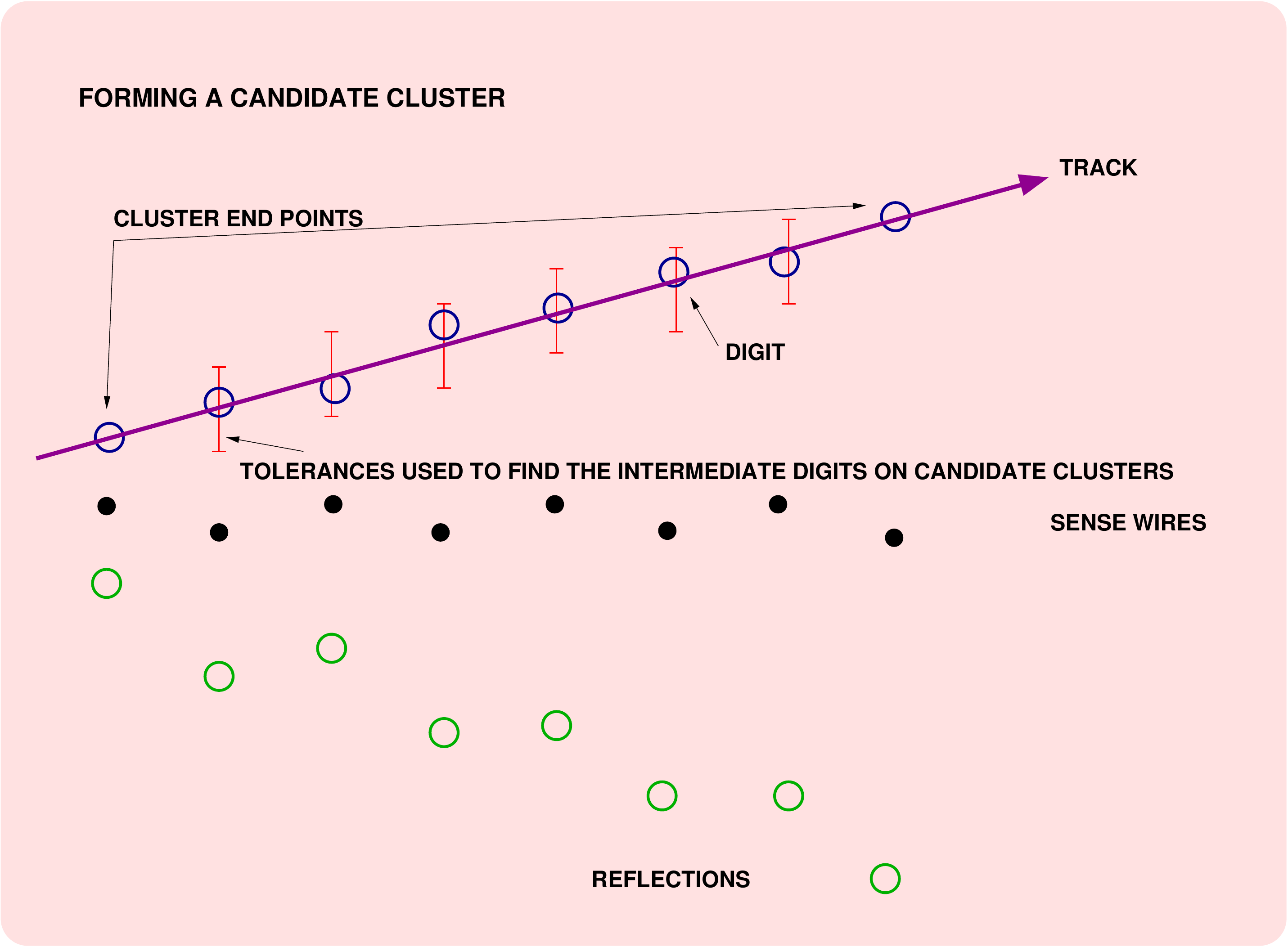}
\includegraphics[width=0.46\columnwidth]{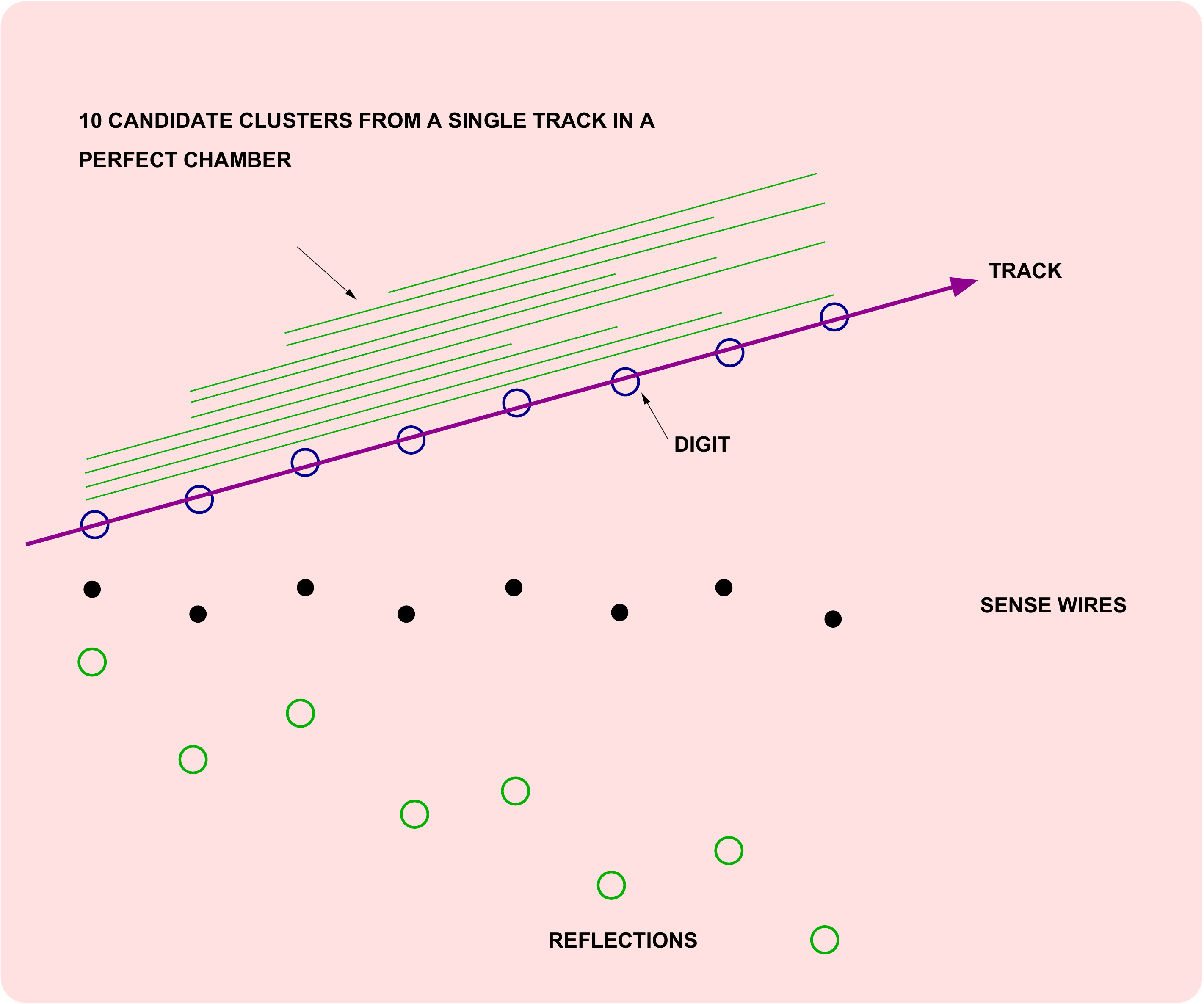}
\caption{(left) Schematic of a candidate cluster in the Q chambers and (right) ten candidate clusters formed from a single track in
  a perfect chamber.}
\label{fig:cluster2}
\end{center}
\end{figure}

The hits identified by the QT algorithm provide sufficient information to define a
plane in the drift coordinate of the module $w$ at the $z$ position of
the module.  The sign of $w$ is ambiguous and thus there are two
hit hypotheses at this stage, but thanks to the stagger in the sense wire positions
the correct sign of $w$ can be identified. If the correct sign is chosen the hits will form a
straight line, whereas the "reflection" hits will be displaced by
twice the sense wire stagger. This is illustrated for a single track
passing through a Q-module with perfect efficiency and no
extraneous pluses in figure $\ref{fig:cluster2}$. 

The pattern recognition of the clusters in the original $4$-wire
P-modules is described in \cite{Burke:1995de}. For those modules, only
clusters containing a minimum of three hits were output as when there
were fewer than three hits the sign of $w$ could not be
determined\footnote{Indeed, the three hit cluster sample contained a
significant proportion of hits with the wrong drift sign and hits from
multiple tracks. }.
The new $8$-wire Q-modules enabled the pattern
recognition to be improved as they have much higher redundancy. In these
modules, except when the number of unusable wires in the module does
not allow it, the minimum number of hits in a cluster is five. Studies
with Monte Carlo showed that the improvements in purity and quality of
clusters above five hits were less significant than below. 

All the possible candidate clusters are formed using all the available
information from all hits. These candidate clusters are then reduced
to a disconnected set of clusters which do not have any hits in
common. The candidate clusters are found by considering the straight
line between any two hits in the same module which are far enough
apart in $z$ that at least three intermediate hits could be found.
Any hits on the intermediate wires that are within three times the hit
resolution of the hypothetical drift distance to the sense wire are
accepted.  Sets of five or more consecutive hits form a candidate cluster.
If a single track is found with perfect efficiency then
this algorithm will generate ten candidate clusters as shown in figure
$\ref{fig:cluster2}$.  This conservative approach leaves the pattern
recognition robust against problems arising from e.g. two tracks in
the same cell, dead or inefficient sense wires and hits originating
from noise.

\subsubsection{Finding a final disconnected set of clusters}

The hits for each candidate cluster are fitted to a straight line at a
fixed $z$ which is the $z$ value of the centre of the module. This
determines the cluster position in a four-dimensional space labelled $(X,
Y, X', Y')$, where $X$ and $Y$ are local analogues to the H1
coordinates and $X'$ and $Y'$ are the slopes in these
coordinates. Each track passing through a Q-module can generate up to
ten candidate clusters which form spikes in this space. 
The candidates in each of
the four dimensions are split wherever the gap between the clusters
exceeds a tuned value for that dimension. 
Candidate cluster spikes with fewer than three candidates are ignored. Each cluster spike is
examined in turn to select the best candidate cluster that it contains. 

Firstly, the candidate cluster with the highest number of unique hits (hits
not shared with clusters in other spikes) is considered. The straight line fit must have a probability of fit of
greater than $5\%$ and this value must be within a factor of ten of
the highest probability for any cluster within the spike. If this is
satisfied, this candidate is chosen and all other candidate clusters
in that spike are eliminated. All remaining spikes which do not have
such a candidate are considered in a second step. The same method for
selecting a single cluster is used but the limit on the fit
probability is reduced to $1\%$.  If this still does not choose a candidate cluster in
a spike, candidates that share clusters with other spikes are removed
and the best candidate in terms of fit probability is taken.

When even a few tracks are incident on the FTD, complex situations can
arise and the candidate clusters which have been chosen from the
spikes can still share hits with other clusters. A very powerful
technique for choosing which of these connected clusters are truly
from tracks and not random combinations is to reduce these clusters to
a final set of disconnected clusters. There is no unique algorithm for
doing this but the method which has been found to work best for the
FTD Q-modules is to remove in turn the cluster which has the highest
number of connections to other clusters. If there is more than one
cluster in this set then for each member of the set, a connection
probability is formed from the following equation:
\begin{equation}
\label{RobProb}
prob\left(\sum_{i=1}^n\chi_i^2,\sum_{i=1}^n\nu_i\right)
\end{equation}
where the sum is over the set of clusters to which this cluster is
connected and $\chi$ and $\nu$ are the chi-squared and number of
degrees of freedom of the connected cluster. The cluster which is
connected to the set of clusters with the highest connection
probability is removed. After each cluster is removed, the number of
connections is recalculated and the disconnection method repeated
until a final, disconnected set of clusters is found.

\subsubsection{Cluster finding efficiency}

Both the cluster finding efficiency and purity need to be considered
simultaneously when judging the performance of the cluster-finding
algorithm. The results in table $\ref{tab:EffPur}$ show the efficiency above
a given purity for $10$, $20$ and $30$ tracks incident on the FTD. 
The efficiency is in general high. However, as the multiplicity increases, the track density is such
that there is significant overlap of the tracks in the projections
measured by the Q-modules. The requirement on purity dictates the size
of the effect on the efficiency.

\begin{table}
\begin{center}
\caption{Efficiency for cluster finding for a given number of tracks in the FTD.
For simulation, the efficiency is given for varying minimum requirements on purity, while
for data the observed efficiency is shown.  The statistical uncertainties are negligible.}
\vspace{0.2cm}
\begin{tabular}{| c | c | c | c | c | }
\hline
Number of tracks in FTD & \multicolumn{3}{|c|}{Efficiency (simulation)} & Efficiency (data)\\
\hline
       & Purity$\>$60\% & Purity $\>$80\% & Purity $\>$95\% & \\
\hline
10 &  98\% & 95\% & 90\% & ~100\%\\
20 &  96\% & 91\% & 88\% & ~100\%\\
30 &  94\% & 87\% & 78\% & ~85\%\\
\hline
\end{tabular}
\label{tab:EffPur}
\end{center}
\end{table}

\subsection{Segments}

The next stage in the pattern recognition software is to identify
track candidates at the level of a supermodule, referred to as
segments. The clusters output from the previous stage are used as
input to a segment-finding algorithm, described in the following.

\subsubsection{Forming candidate segments}
\label{subsubsec:segmentform}

Clusters define planes which contain the trajectory of a charged
particle. The modules are rotated in $\phi$ with respect to
$\phi=0^{\circ}$, such that modules have a unique orientation
within a supermodule (see figure $\ref{Fig:FTDorientations}$). The
intersection of clusters from different modules define segments.  At
high multiplicities, e.g. more than $20$ tracks, the combinatorics
means that intersections of only two clusters have very low
purity. Therefore, a segment is defined to be the intersection of
three or more clusters. All intersections between any two clusters in a
supermodule are found and fitted to a straight line at a fixed $z$ (the P-module
wire closest in $z$ to the Q modules). In this way, each
intersection is represented by a point in $(X, Y, X', Y')$ space. If a
track is perfectly measured and found in all five modules within a
supermodule, then there will be ten intersections all within close
proximity in the four dimensional space.
The existence of three or more lines of intersection is considered a candidate segment, in order to account
for detector effects. 

Candidate segments are identified as spikes in
the four dimensional intersection fit space $(X, Y, X', Y')$ and are
found in a similar way to the cluster spikes described previously. The
fit results from all of the two cluster intersections are ordered in
each dimension. A scan is made for a gap above a given threshold,
which is tuned for each dimension using Monte Carlo and corresponds to
several times the detector resolution in that dimension. Candidate
segments are three or more intersections within a given intersection
spike. For each of the candidate segments, all of the hits belonging
to each of its clusters are fitted to a straight line. In order to
identify true track segments, these candidate segments are fed into an
algorithm which finds a disconnected set using a connectivity
table. The algorithm developed for disconnecting the set of connected
candidate segments looks for any singly connected segments and removes
them. If there is more than one such candidate, the one connected with
the highest chi-squared probability is removed first. The connection
probability is again defined as in equation $\ref{RobProb}$, where now
the sum is over the chi-square and degrees of freedom of the connected
segments. If no singly connected segments exist then the candidate
segment which is connected with the highest connection probability is
removed, consistent with it being in the set of segments with the
highest number of connections to other segments. The algorithm is
repeatedly applied after a candidate segment is removed until a
disconnected set remains.

\subsubsection{Segment forming efficiency}

\begin{figure}
\begin{center}
\includegraphics[width=0.5\columnwidth,height=0.45\textheight]{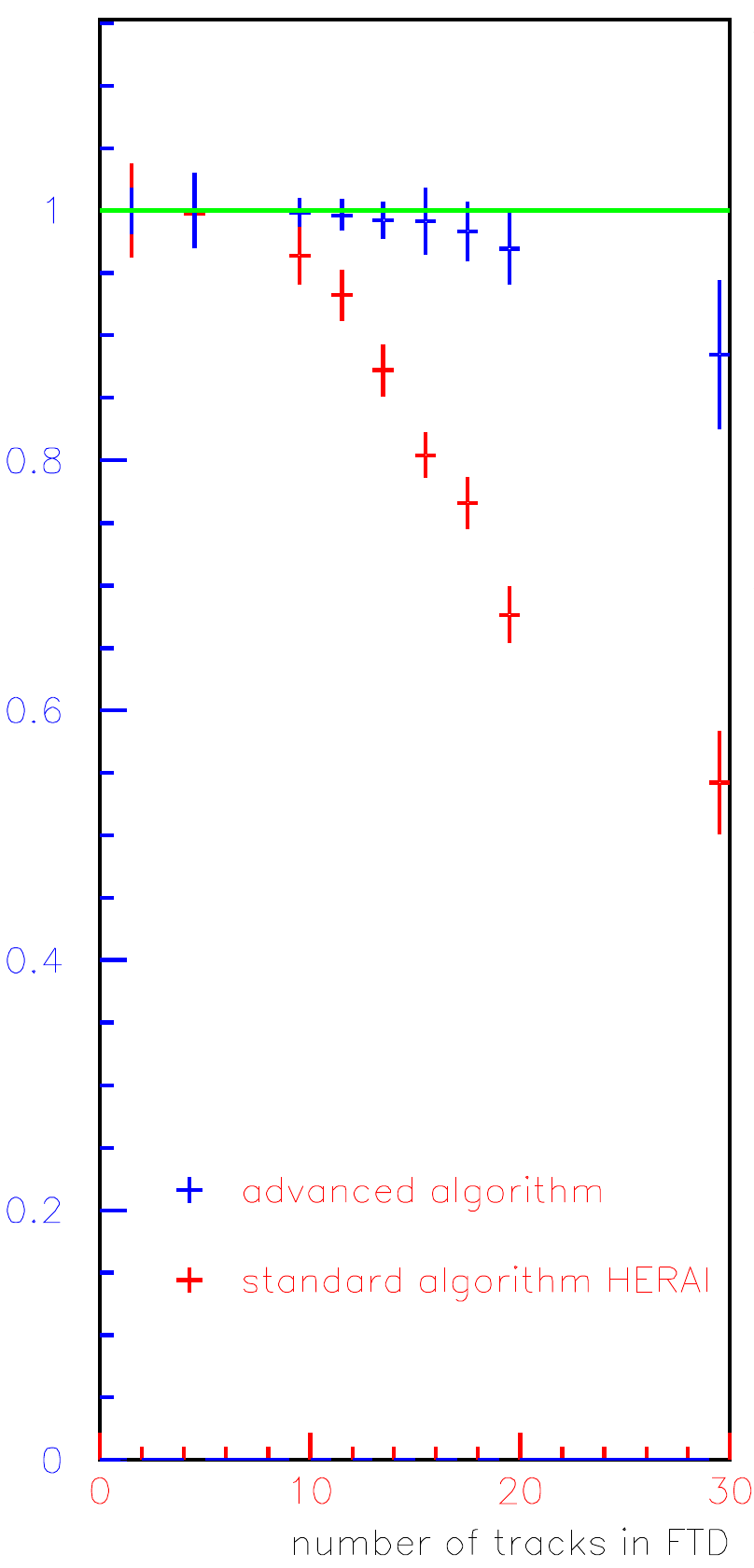}
\caption{Figure showing the efficiency of segment finding in the
FTD as a function of track multiplicity, comparing the old detector and software (red) with the
new detector and software (blue).}
\label{fig:perf}
\end{center}
\end{figure}

The efficiency for finding segments is shown in figure
$\ref{fig:perf}$ as a function of the number of tracks in the FTD. The
plot shows the segment finding efficiency for tracks which
have a purity greater than $60\%$ (the great majority of tracks have a
purity of nearly $100\%$). The red points show the efficiency for the
old FTD with the old software algorithms. The blue points show the
same for the new FTD and the new algorithms. A large improvement
at high multiplicities is clearly demonstrated with an efficiency of
$100\%$ at low multiplicities, remaining around $90\%$ for an event
with $30$ tracks in the FTD.  For context, the number of segment candidates
found increases with increasing track density, such that the algorithm for
finding the disconnected set of segments is crucial. At twenty tracks, $\sim75\%$ of
the candidate segments must be accurately removed in order to uncover
the true tracks. This is achieved with such accuracy that an
efficiency of approximately $97\%$ is obtained. It is clear that the
upgraded FTD, together with this pattern recognition is a great
improvement on the previous FTD and has good efficiencies for most
topologies. The segments from this stage are then passed to a linking
stage to make them into tracks, which uses the same algorithm as 
described previously \cite{Burke:1995de}.

\section{Improvements in the Pattern Recognition using a Kalman Filter}

\subsection{Introduction}

Essentially the same Kalman Filter framework as described
in \cite{Burke:1995de} is used to determine the optimum track parameters for each
track in the FTD, albeit with two major additional modifications
(other than those necessary to treat the new Planar chambers rather
than the old Radial chambers). These modifications are to add pattern
recognised track clusters to a track during the Kalman filtering and
then to selectively reject track clusters from a track should it fail
to be accepted after the Kalman filter, until the resultant track is
acceptable. Both of these are techniques for adding or removing
pattern recognised objects to or from the track which is otherwise
processed on a point by point basis as the filter moves from
measurement surface to measurement surface. The first modification compensates for
inefficiencies in the detector or the pattern recognition within a
single supermodule such that unused clusters may be attached to a
track found in the other supermodules and the second allows pattern
recognition mistakes to be removed from a track. 
These two modifications are described separately below.

\subsection{Cluster addition}

The original implementation of the Kalman Filter in the FTD already
included the option to add hits during both the filtering and
smoothing steps. Unfortunately the lack of constraint on individual hits
meant that this was not particularly useful and the code was never
used in production. The modification was to use this existing code, but with a
preselection as to which hits may be considered for addition. The Kalman
Filter steps from one measurement surface to the next, first from the
interaction point outwards (filtering) and then back (smoothing). At
the transition between modules or orientations the preselection checks
if there are no hits on the track in the module just entered. If so,
it then projects the incoming track state vector through each
measurement surface in the module and searches to see if there is a
nearby unambiguous unused candidate cluster in this module, respecting
the resolved left/right ambiguity in the pattern recognised
cluster. If so, at each measurement surface within the module it then
uses the existing pointwise code to consider whether to add the point from
this cluster, again respecting the resolved left/right
ambiguity. If on any pass through the FTD a hit is added the
filter then iterates until there has been a complete filter and
smooth pass without adding any points. Finally, should the filter fail
for any reason during this process the code reverts to the original
implementation without attempting to add any clusters. 

\subsection{Cluster rejection}

Again, the original implementation of the Kalman Filter in the FTD
already included the option to reject hits during both the filtering
and smoothing steps. However the covariance matrix proved poorly
behaved when removing the contribution from the point being rejected,
even when using double precision for all arithmetic. An alternative
prescription was therefore adopted. The Kalman filter was run to
completion without point rejection. The resultant track quality was
then examined and if acceptable no further action taken. If however,
the track would then be rejected, it was rather reprocessed, using the
final state vector as input but removing the cluster contributing most
to the overall errors, iterating until either the resultant refiltered
track was acceptable or too few clusters remained on the track. 

\subsection{Results}

\begin{figure}
\begin{center}
\includegraphics[width=0.9\columnwidth]{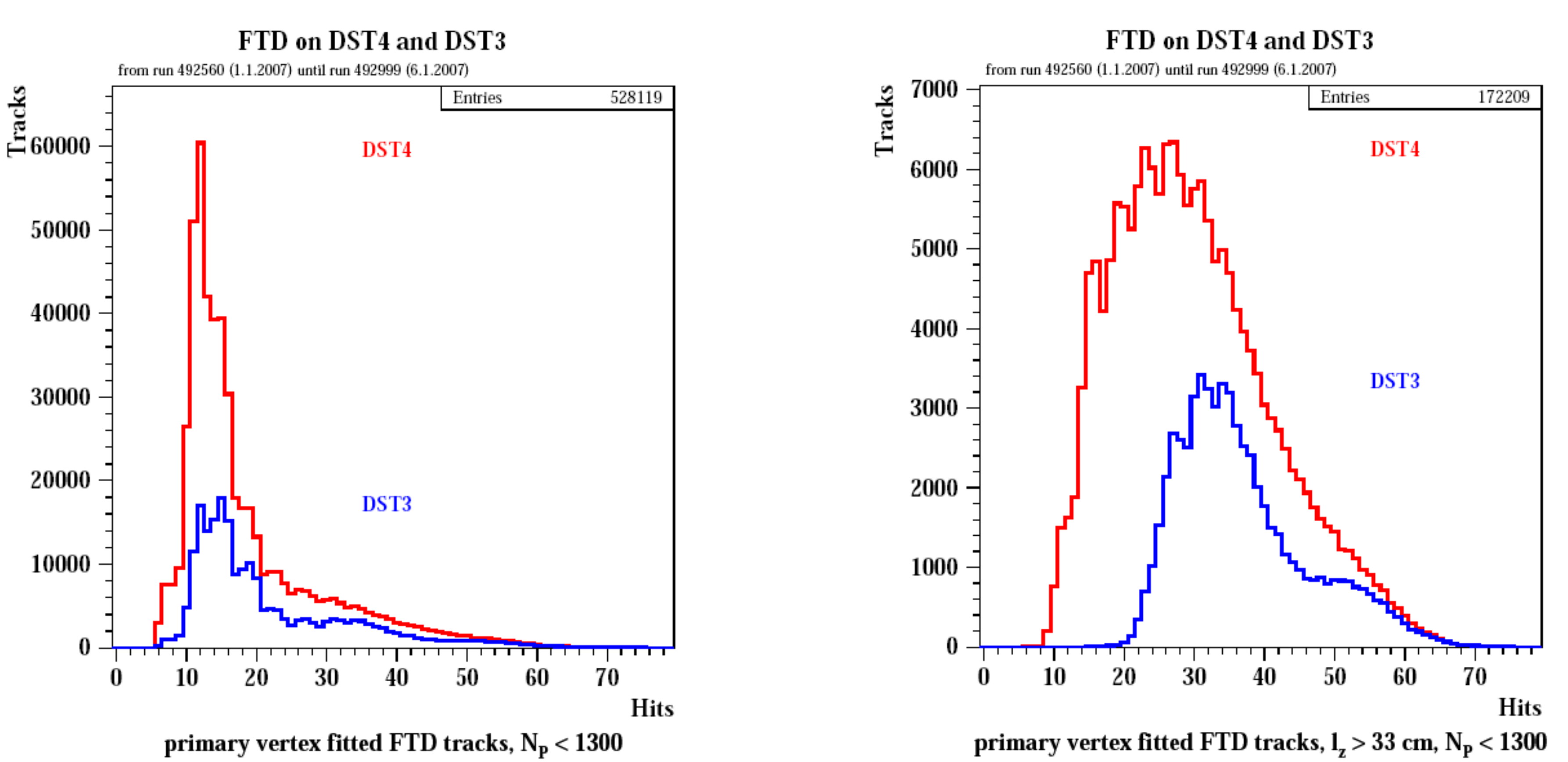}
\caption{The number of hits on
tracks found in the original implementation (DST3) and with the
cluster addition and rejection modifications (DST4) for all Forward
tracks (left) and for Forward tracks passing through more than one
supermodule (right).}
\label{Fig:Kalmandst3-4comparison}
\end{center}
\end{figure}

The effect of these modifications is clearly visible in figure
$\ref{Fig:Kalmandst3-4comparison}$ which shows the number of hits on
tracks without the Kalman filter modifications (DST3) and with them (DST4).
The structure arising from tracks being found
in one, two or all three modules is clearly visible in the former,
whereas the in-fill from the first modification and outlier rejection
from the second result in the distribution of the number of hits on
tracks in the end result being much smoother, suggestive of the resultant
tracks being pattern recognised in the FTD as a whole.

\section{Combining FTD and CTD Information}

\subsection{Vertex fitted tracks}

The FTD constructs a vertex candidate, with moderate precision, from
FTD tracks using a weighted average of the $z$-information from each
track. A superior vertex is formed from the CTD, which incorporates
information from all of the tracking detector technologies in the
central region of the H1 detector. This vertex candidate has very high
precision, thanks largely to the silicon detectors located close to
the H1 beam pipe\cite{H1Silicon}. Only in the rare cases where the CTD fails to
find a vertex candidate is the FTD vertex candidate used. The
track-to-vertex fitting process simply adds the vertex as a spacepoint in a Kalman
Filter fit of the forward track, yielding improved parameters for that
track.  The resultant track is referred to herein as a "Forward" track.

\subsection{Combined FTD and CTD tracks}

Linking of tracks reconstructed using the CTD ("Central") and FTD to form "Combined"
tracks is performed in order to produce tracks which, in principle,
have more precise estimates of the true track parameters (assuming
that the correct associations are made). Tracks from each
device are extrapolated to the inner surface of the CTD end-wall, with track
parameters and their associated measurement errors, including multiple
Coulomb scattering contributions using the Kalman Filter.
The two sets of track parameters are then compared,
potential links identified and ambiguities resolved. The parameters of
surviving matched track pairs are combined using a weighted
average.

The inputs to the the combination algorithm are
the lists of Central and Forward tracks. The two
track types use two different parameterisations to represent the
track, shown in table $\ref{TAB:TrackParams}$.

\begin{table}
\begin{center}
\caption{Track parameterisation types used for Central and Forward tracks.}
\vspace{0.2cm}
\begin{tabular}{|c|c|c|}
\hline
   & Type 1 (Central)& Type 2 (Forward)\\
\hline \hline
  Curvature & $1/r$& $1/r$\\ 
  Azimuthal angle &$\phi$& $\phi$\\
  Polar angle &$\theta$& $\theta$\\
  Position 1 &$dca$& $x$\\
  Position 2 &$z0$& $y$\\
  Location &-& $z$\\
\hline
\end{tabular}
\label{TAB:TrackParams}
\end{center}
\end{table}

These are different due to the orthogonal sense-wire geometries of the
two trackers; tracks in an intermediate range of polar angles (between
approximately $20$ and $70$ degrees) can be represented in either form,
but in general they are not interchangeable. Note that the $z$
coordinate in Type 2 is not a fit parameter (i.e.it has no error); it
just specifies where on the track the other parameters are given. The
equivalent parameter for Type 1 is implicit, as the parameters are
always given at the point of closest approach to the $z$ axis.

Central and Forward tracks are extrapolated to the common $z$-plane
(the CTD end-wall) using the Type 2 parameterisation.
The Forward track parameters are first corrected for any misalignment
between the Forward and Central trackers, using pre-determined
translation and rotation adjustments (see section $\ref{sec:align}$). 
Multiple scattering is allowed for by increasing the
extrapolated covariance matrix, assuming material with known radiation
length between planes either side of the CTD end-wall. Energy loss is
corrected using a Bethe-Bloch parameterisation and the same 
approximate model for material, with the assumption that the track is a pion.

A list of (possibly ambiguous) associations between Forward and
Central tracks is created by looping over all extrapolated Central
tracks, and for each of these calculating a chi-squared with each
Forward track, using the extrapolated track parameters. Candidate
associations with an acceptable chi-squared are kept.
All Forward tracks with only one associated Central track are accepted
as a Combined track candidate.
Ambiguities are resolved and links chosen by looping over all Forward
tracks and choosing the association with the minimum
chi-squared\footnote{If there are multiple associations, Central tracks
with a larger number of hits are preferred even if they have a larger
chi-squared than an association with a track having fewer hits. This
means that Central tracks with $z$-chamber hits are preferred as such
tracks have a superior $z$-resolution.}. 
After each link is made, any other associations with the
same Central track are deleted, so that the final list of links is
unambiguous. Finally, the Combined tracks themselves are created by taking the weighted mean of
the two sets of track parameters and are then transformed to the Type 1 parameterisation.

\subsection{Alignment to the CTD}
\label{sec:align}

\begin{figure}[h]
  \centering
  \includegraphics[height=0.25\textheight, width=0.6\linewidth]{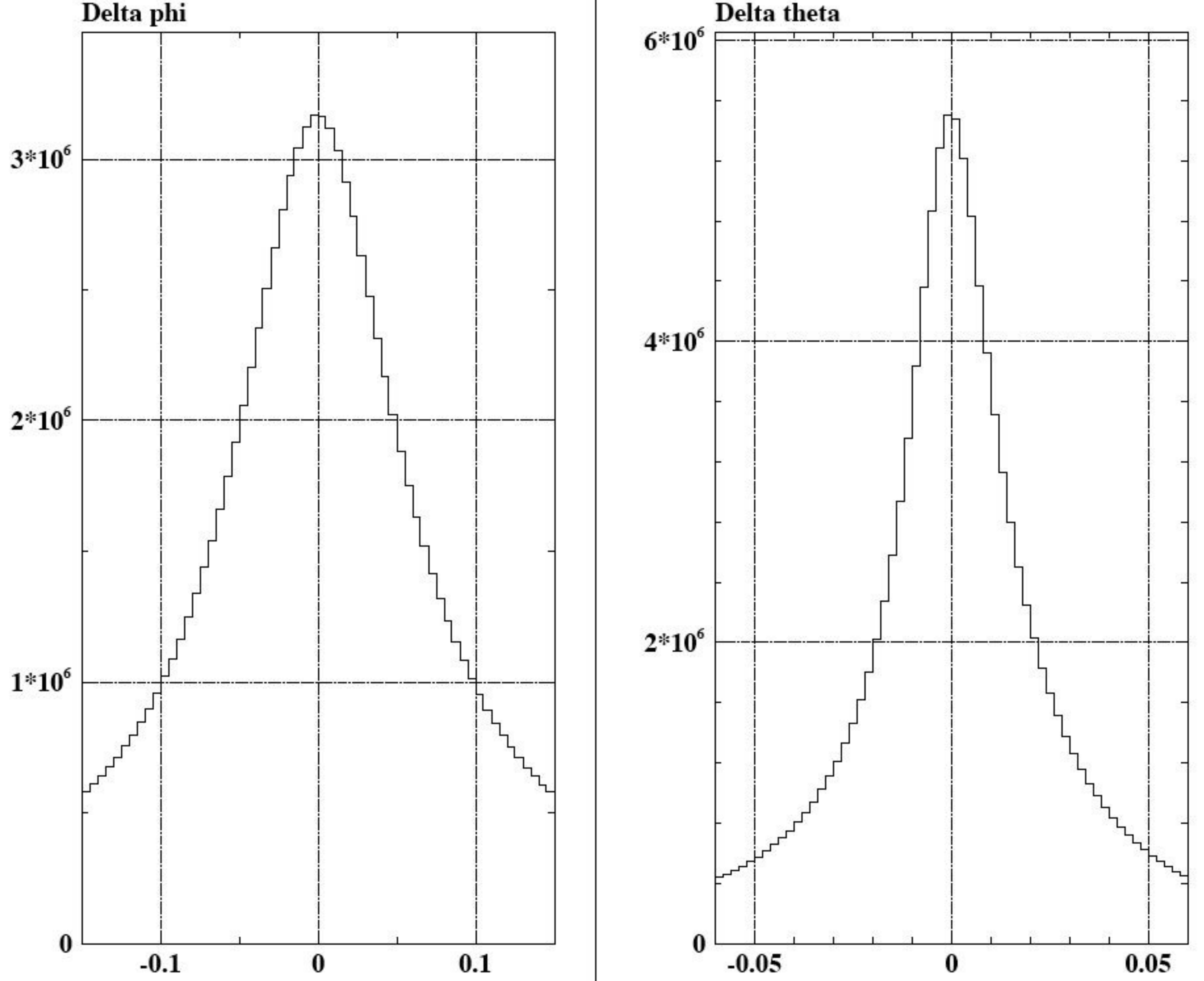}\\
  \vspace{0.5cm}
  \includegraphics[height=0.25\textheight, width=0.6\linewidth]{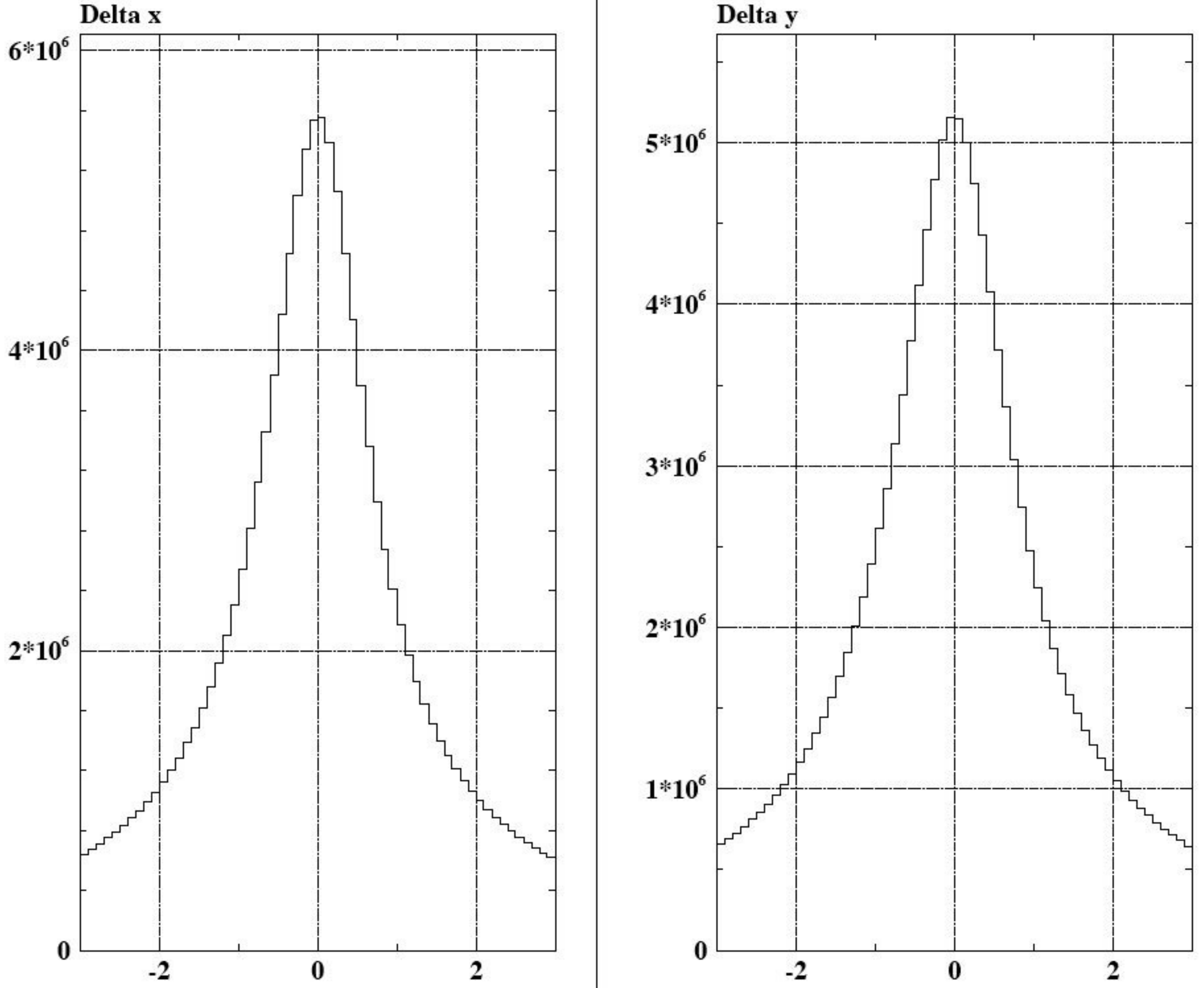}
  \caption{(top) Delta phi and Delta theta (radians) and (bottom) Delta x and Delta y (cm)
  between Forward and Central tracks at the CTD end-wall.}
  \label{fig:ktnrec}
\end{figure}

The residuals measured at the CTD end-wall in the track-linking procedure, 
in $x$, $y$, $\theta$ and $\phi$, provide important constraints on the relative
alignment of the FTD and CTD, they are shown in figure $\ref{fig:ktnrec}$. 
By convention H1 chooses the CTD as its
reference system, with respect to which all other sub-detectors must
be aligned. Regarding the FTD as a rigid body, there are six degrees
of freedom that determine its relative alignment with respect to the
CTD; a three-vector translation to align its coordinate origin with
that of the CTD, plus three Euler rotations that define the relative
angular orientation of the two systems. Thus, the residuals monitored
in the track linking provide half of the six required alignment
constraints (the $\phi$ residual effectively measures one of the
three Euler rotations). The other three constraints can be derived
from the track-to-vertex fitting process described previously, with
the residuals in $x$, $y$ and $z$ with respect to the vertex providing the information.
The final set of six alignment parameters (a three-vector translation plus three Euler
rotations) are used to align Forward tracks to the rest of H1. The
equations relating these six parameters to the six observed residuals are
non-linear and are solved numerically, using an iterative procedure to
derive the parameter set that minimise (to zero) the observed six alignment residuals. 

\section{Performance of the HERA II FTD}

The performance of the FTD has been assessed using
Monte Carlo simulation and real data. The simulation of the FTD is
described in \cite{Burke:1995de}. It should be noted that, for HERA II, the
track-related noise simulation and parameterisations of the pulse
shape fluctuations are not used. However, the efficiencies of the
detector were tuned to match those seen in data. In the following,
Forward tracks are studied, with a vertex which is always defined by the CTD.
The tracks must pass a selection, detailed in section $\ref{sec:tracksel}$,
and the resulting FTD performance is monitored using independent H1
detectors. Data is compared to the simulation for both an inclusive
sample of neutral current (NC) deep-inelastic scattering (DIS) events and an elastic $J/\Psi$
sample of events, representing high and low multiplicities,
respectively. 

\subsection{Track selection}
\label{sec:tracksel}

The surroundings of the FTD further
complicates track reconstruction in an already demanding
environment. The readout electronics of the CTD in the CTD end-wall are
situated directly between the CTD and the FTD and constitute
approximately $X_0/2$ of dead material. This dominates the multiple
scattering component of the FTD resolution and makes low momenta
inaccessible. A collimator is situated directly underneath the FTD and
generates many extra tracks from secondary interactions with the proton
remnant. Therefore, only a portion of the tracks which are
reconstructed by the FTD are useful for physics analysis and a
selection is applied to Forward and Combined tracks\footnote{Studies
of track parameters and their errors showed that the track
selection also improved the scaled residuals for track parameters.}. 
Tracks passing this selection are used in the HFS algorithm of
H1. The selection is given in table $\ref{tab:selection}$.

\begin{table}[h]
  \centering
  \caption{The cuts used to select Forward and Combined tracks for physics analyses.}
\vspace{0.2cm}
  \begin{tabular}{|l|l|l|}
    \hline
     Cut motivation & Forward & Combined \\ \hline \hline
     Fiducial & $6^{\circ} < \theta < 25^{\circ}$ & $10^{\circ} < \theta < 30^{\circ}$ \\
     & RadialStart $> 25$ cm & RadialStart $< 50$ cm \\ \hline
     Pointing & $R0 < 20$ cm & ${\rm DCA^{'}} < 5$ cm \\ \hline
     FTD-only & NHits $> 10$ &  \\ 
     & $z_e - z_s > 10$ cm & \\ \hline
     Reliable measurement & $p > 0.5$ GeV & $p > 0.5$ GeV \\
     & $p_T > 0.001$ GeV & $p_T > 0.12$ GeV \\ \hline
     Outlier removal & $dp/p < 9999.9$ & $dp/p < 9999.9$ \\
     & $\chi^2({\rm VF}) <25$ & $\chi^2({\rm VF}) < 50$ \\
     & $\chi^2({\rm Track}) < 10$ & $\chi^2({\rm Link}) < 50$ \\
    \hline
  \end{tabular}
  \label{tab:selection}
\end{table}

The selection can be broken down into three
components. Fiducial cuts are applied to ensure that the track
originated from an $ep$ collision close to the nominal interaction
point. The requirement of a minimum starting radius of a Forward
track avoids the areas of the detector most affected by ageing, while
the maximum starting radius for Combined tracks ensures good overlap
with the CTD. FTD tracks are required to have a minimum length and
number of hits to ensure good detector acceptance and avoid problems
due to hardware failures (Combined tracks have
an inherently good acceptance due to the additional CTD
information). A minimum momentum cut of $0.5$~GeV is applied in order to have
good acceptance (the particle should have enough energy to get through
the CTD end-wall). The minimum transverse momentum cut for Combined
tracks is applied to ensure that the central track component is well
measured. Finally, tracks with very poor momentum measurements and
outliers in the $\chi^2$ distributions are removed as such tracks are not well
simulated. 

\subsection{Comparisons of data and simulation for inclusive NC events}

Inclusive DIS events at high $Q^2$ and $x$ produce large numbers of
particles in the forward region from the struck parton. In principle,
the kinematics of NC DIS events can be reconstructed
solely from the kinematics of the scattered electron, but an improved
resolution is achieved if the kinematics of the hadronic final
state (HFS) are also included. In the case of charged current DIS
events, the kinematics must be measured from the HFS
alone. In the following, a standard inclusive NC DIS
selection is used to select a sample of events which are globally very well
described by simulation.

\begin{figure}[h]
  \centering
  \includegraphics[width=0.49\linewidth]{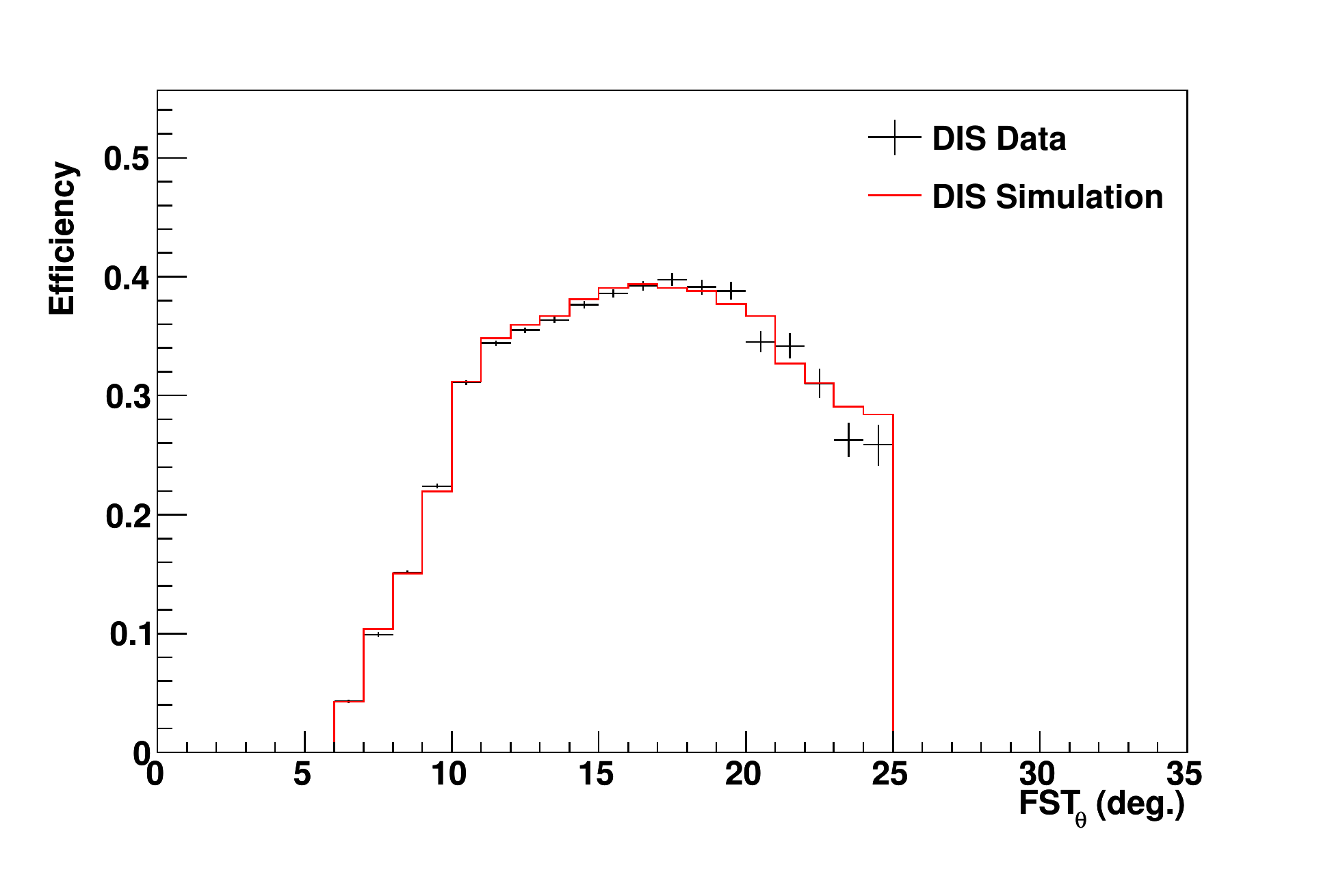}
  \includegraphics[width=0.49\linewidth]{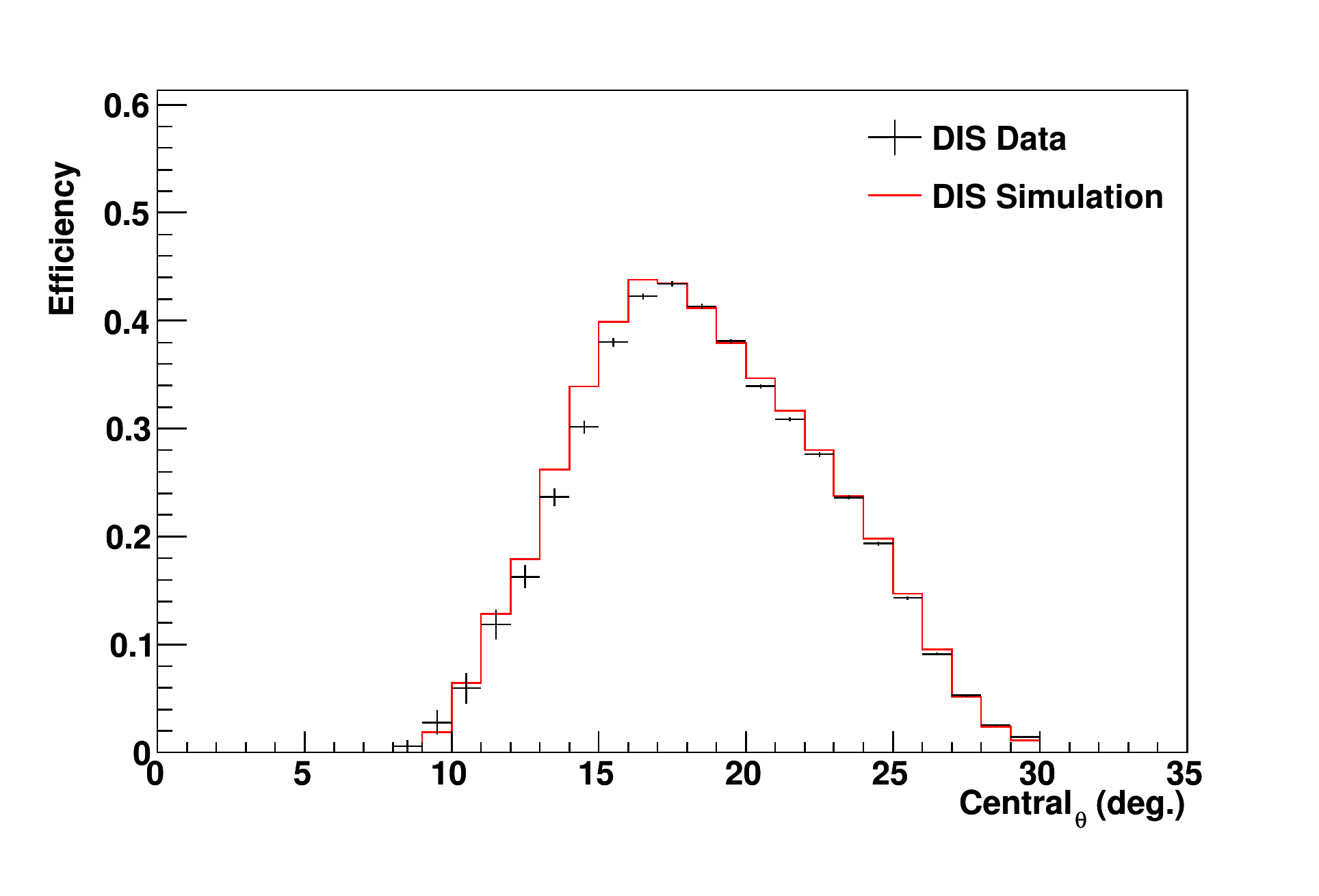}
  \includegraphics[width=0.49\linewidth]{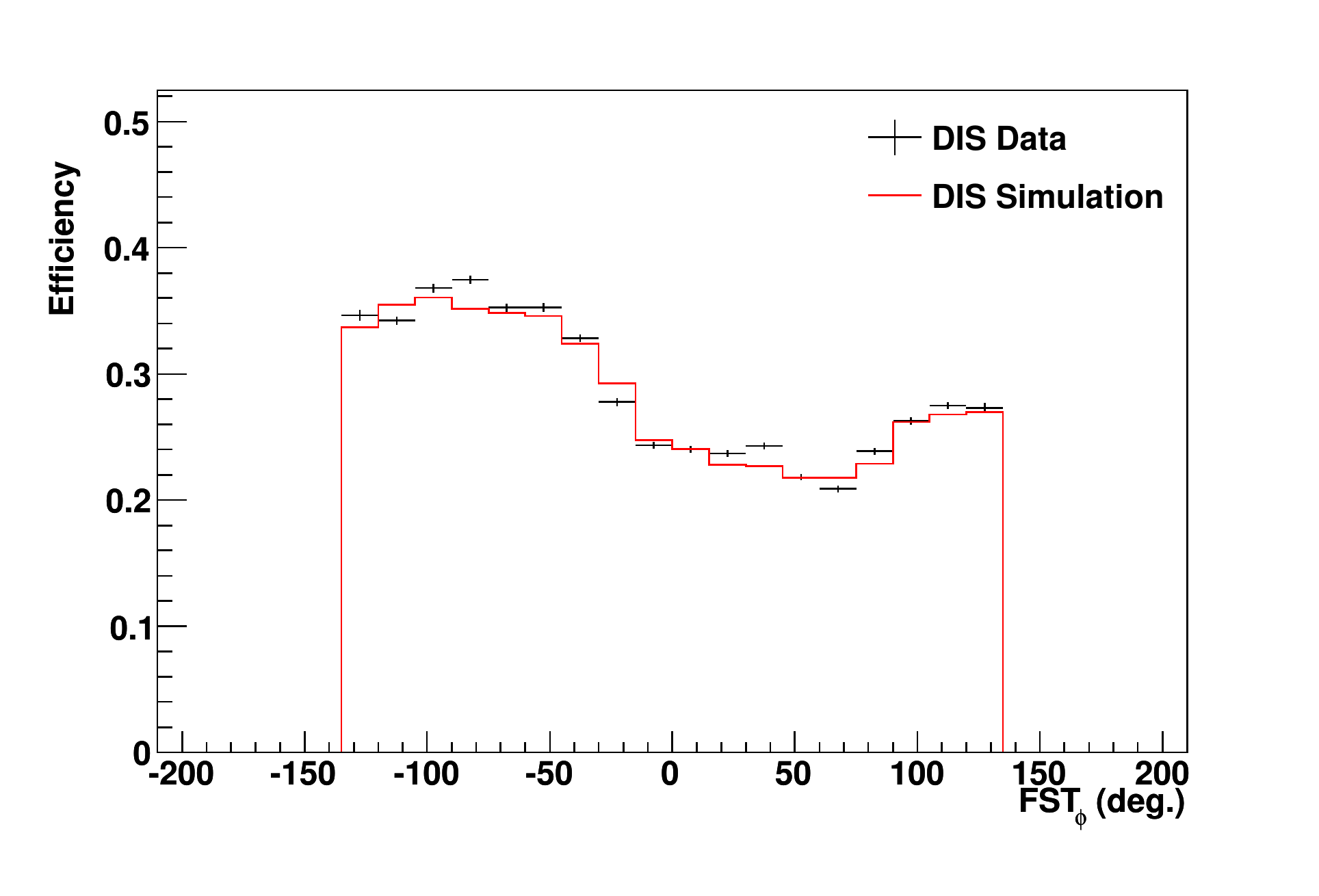}
  \includegraphics[width=0.49\linewidth]{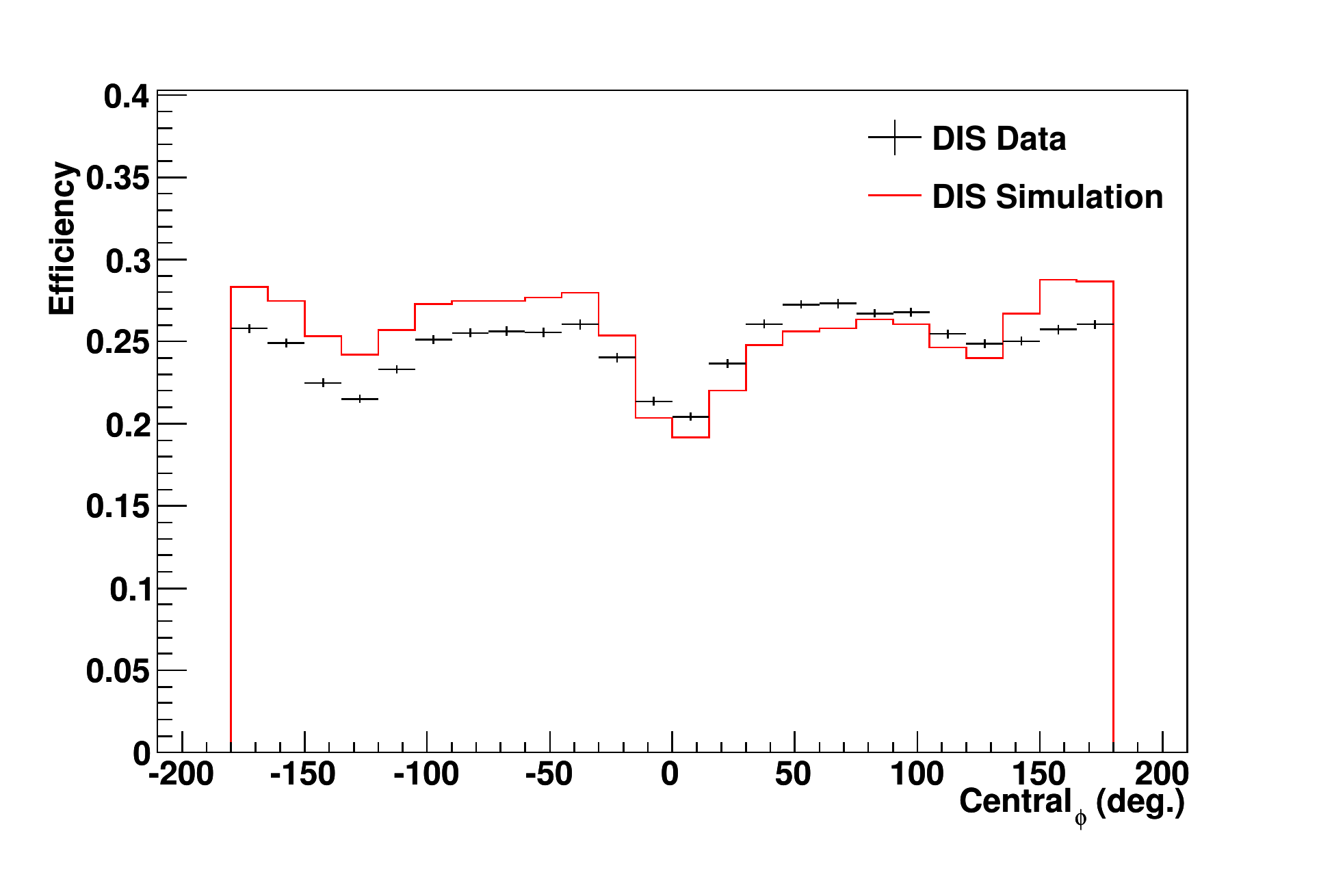}
  \includegraphics[width=0.49\linewidth]{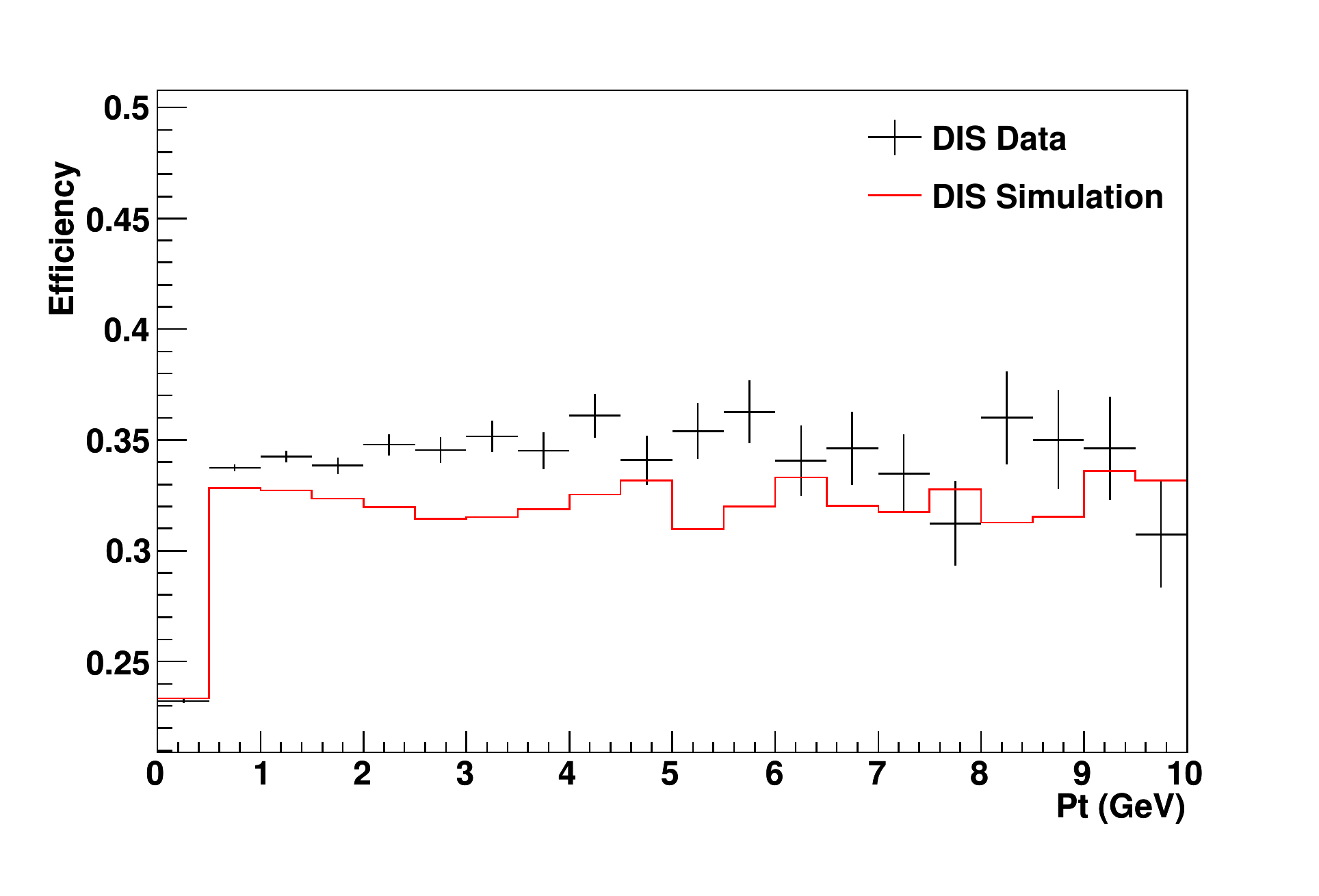}
  \includegraphics[width=0.49\linewidth]{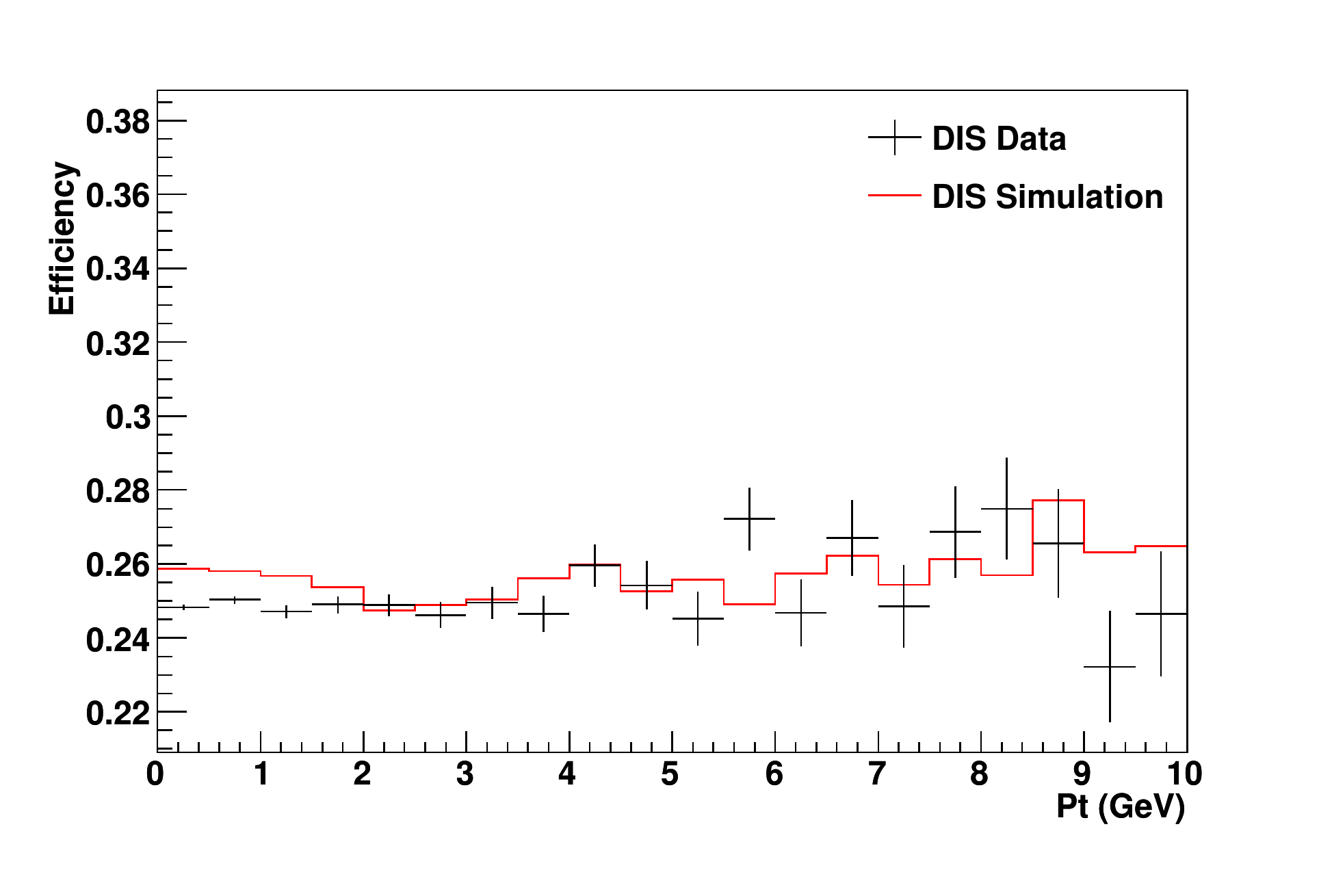}
  \includegraphics[width=0.49\linewidth]{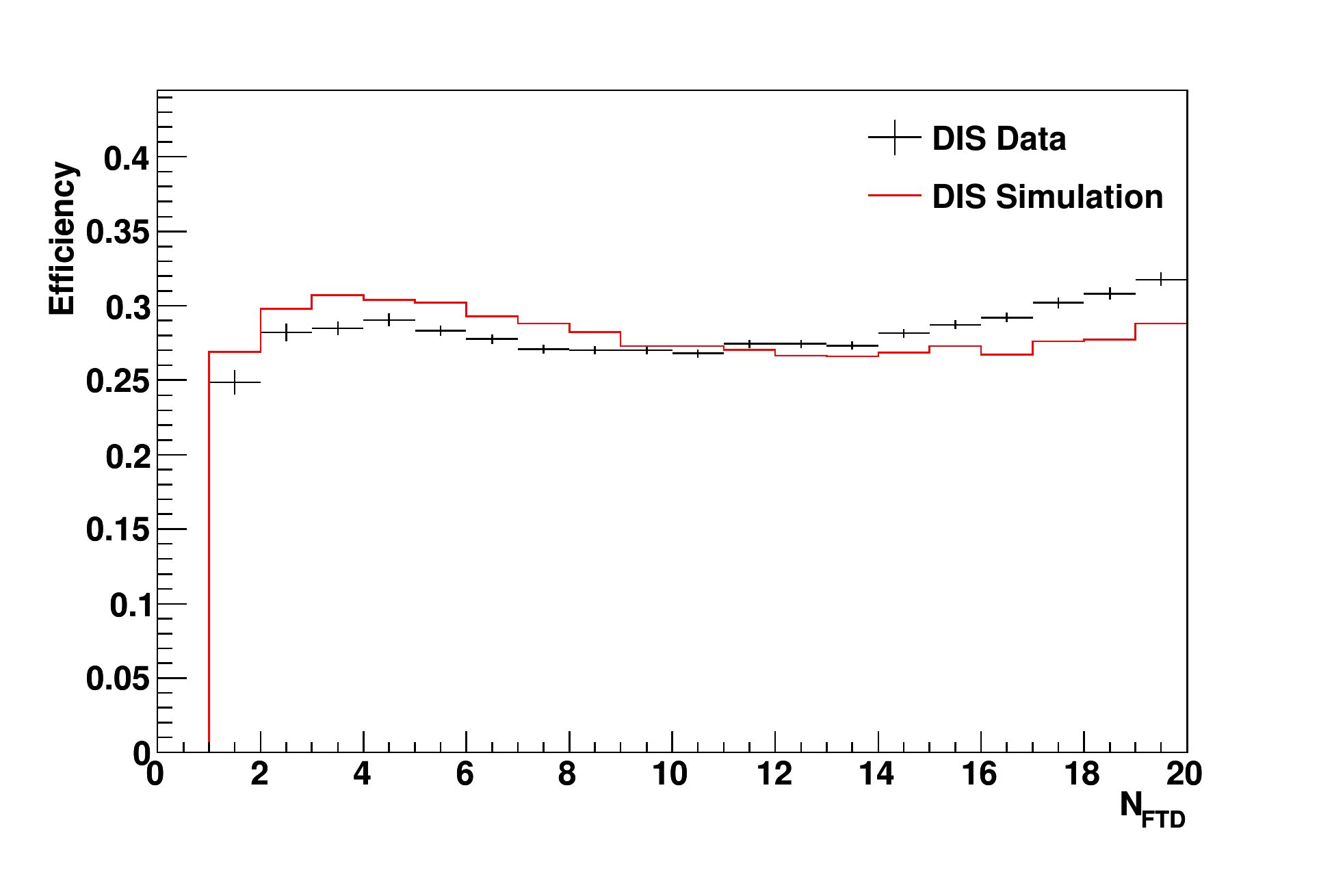}
  \includegraphics[width=0.49\linewidth]{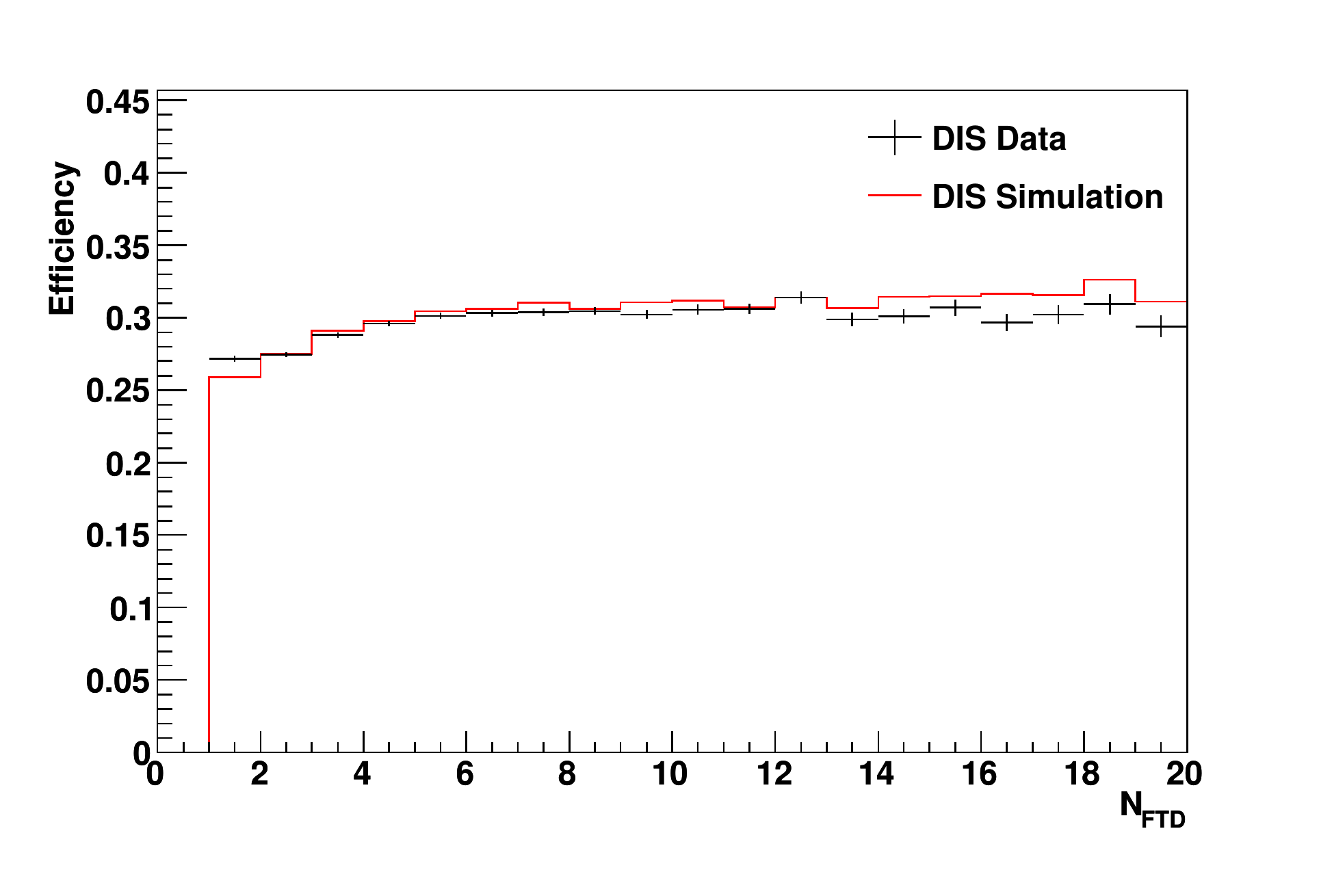}
  \caption{The efficiency of finding a matched Forward (Combined) track as measured
  by the FST (CTD) as a function of $\theta$, $\phi$, $p_T$ and FTD track multiplicity.
  Forward tracks are studied in the left column, Combined tracks in the right.}
  \label{fig:incperf}
\end{figure}

\begin{figure}[h]
  \centering
  \includegraphics[width=0.49\linewidth]{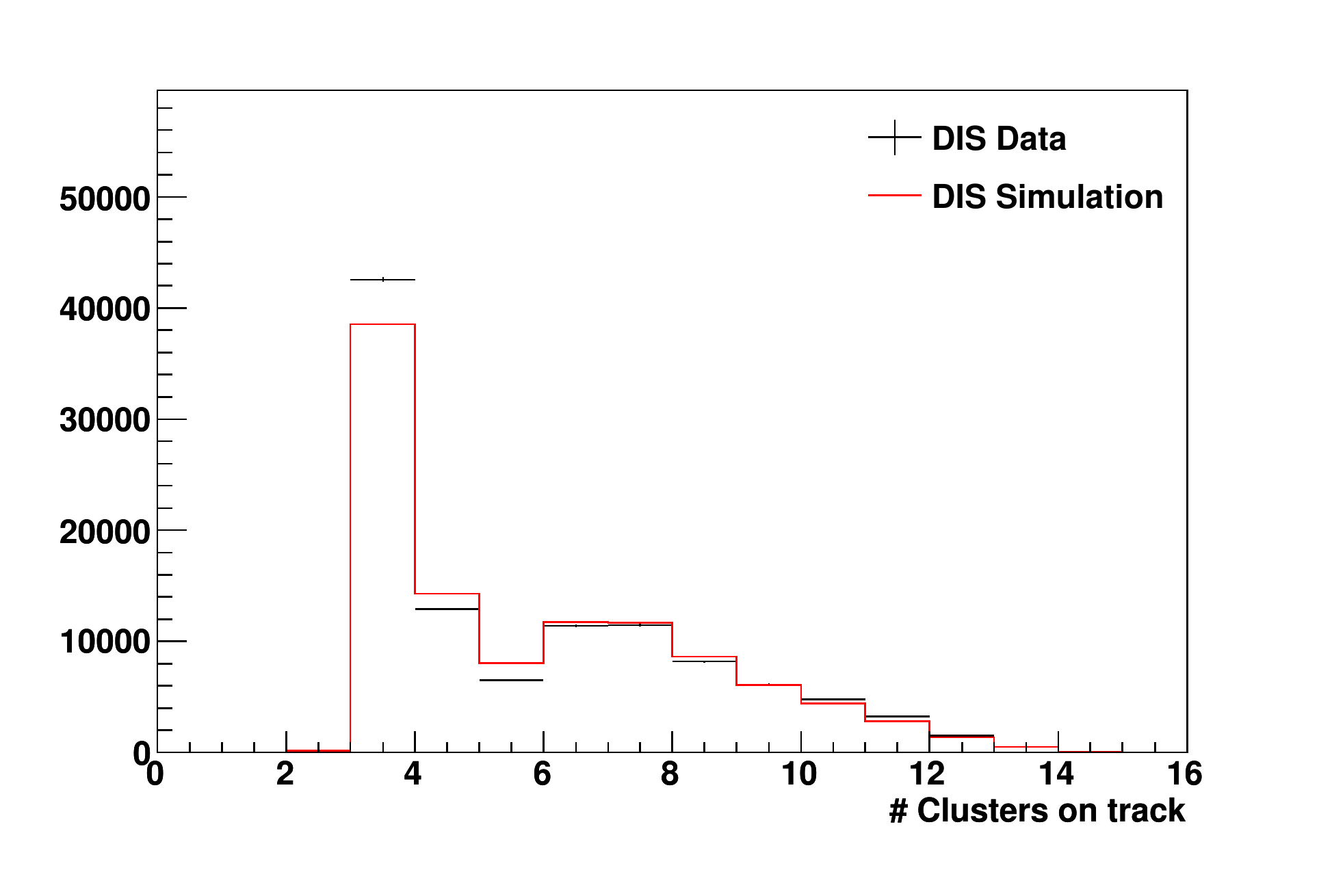}
  \includegraphics[width=0.49\linewidth]{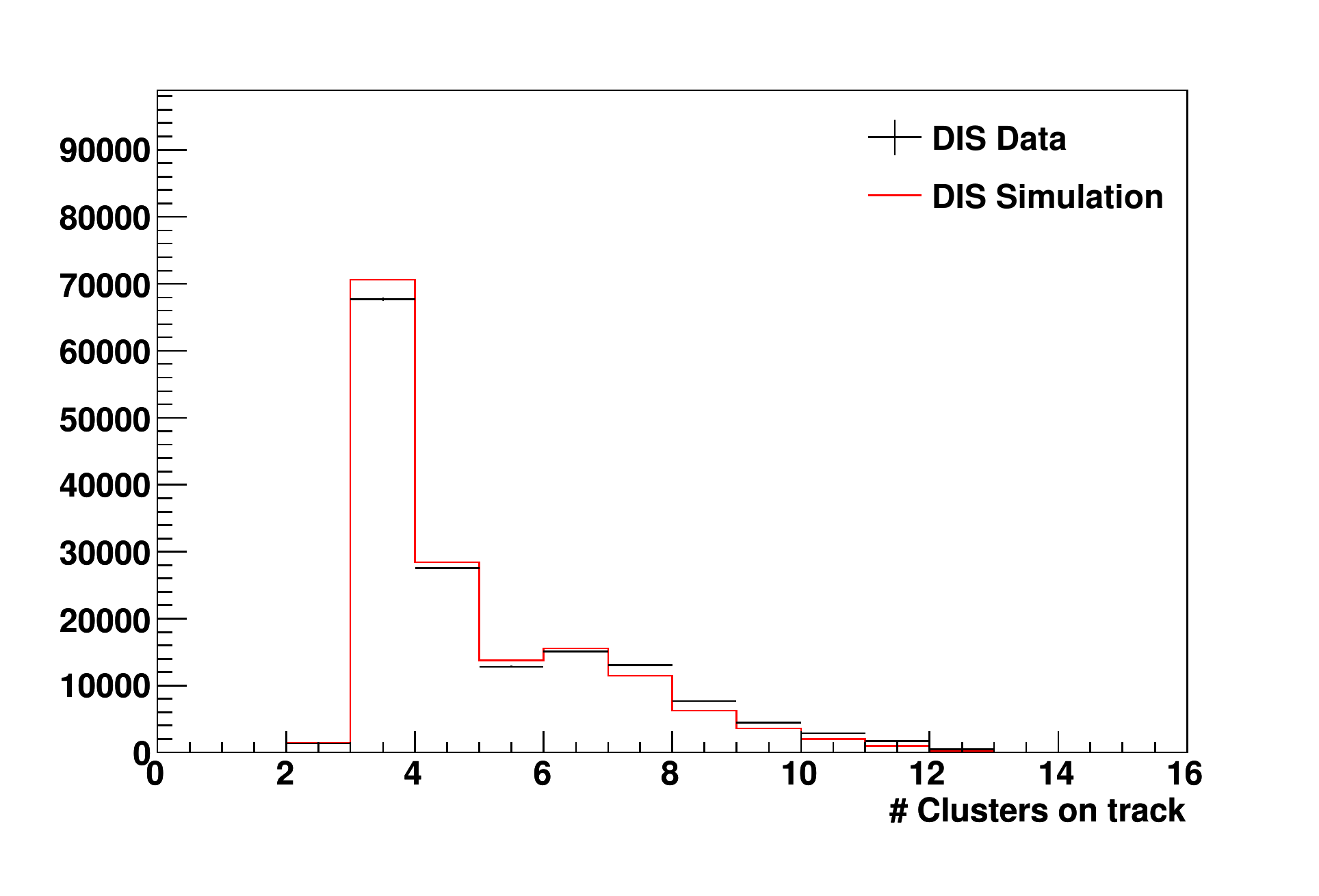}
  \includegraphics[width=0.49\linewidth]{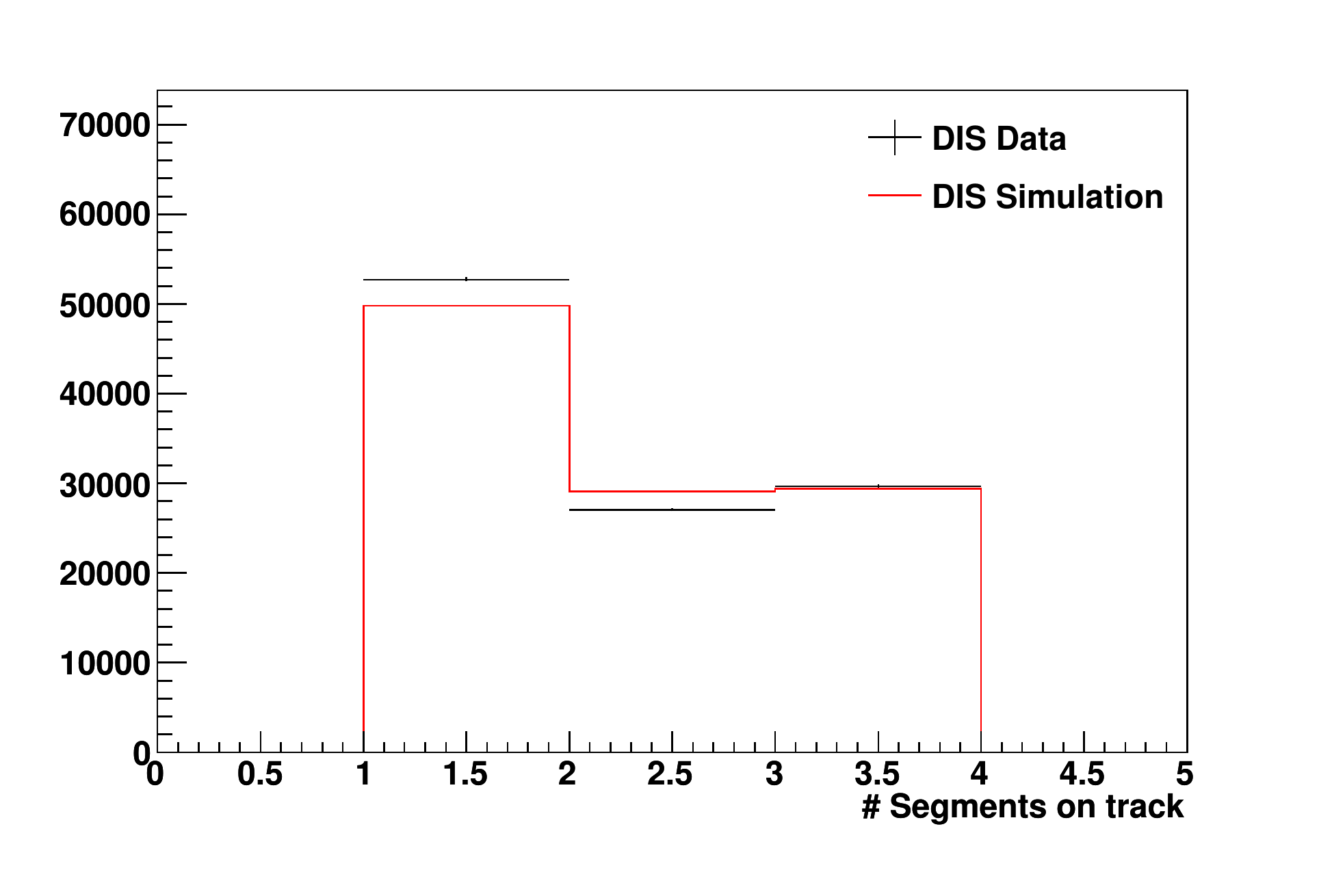}
  \includegraphics[width=0.49\linewidth]{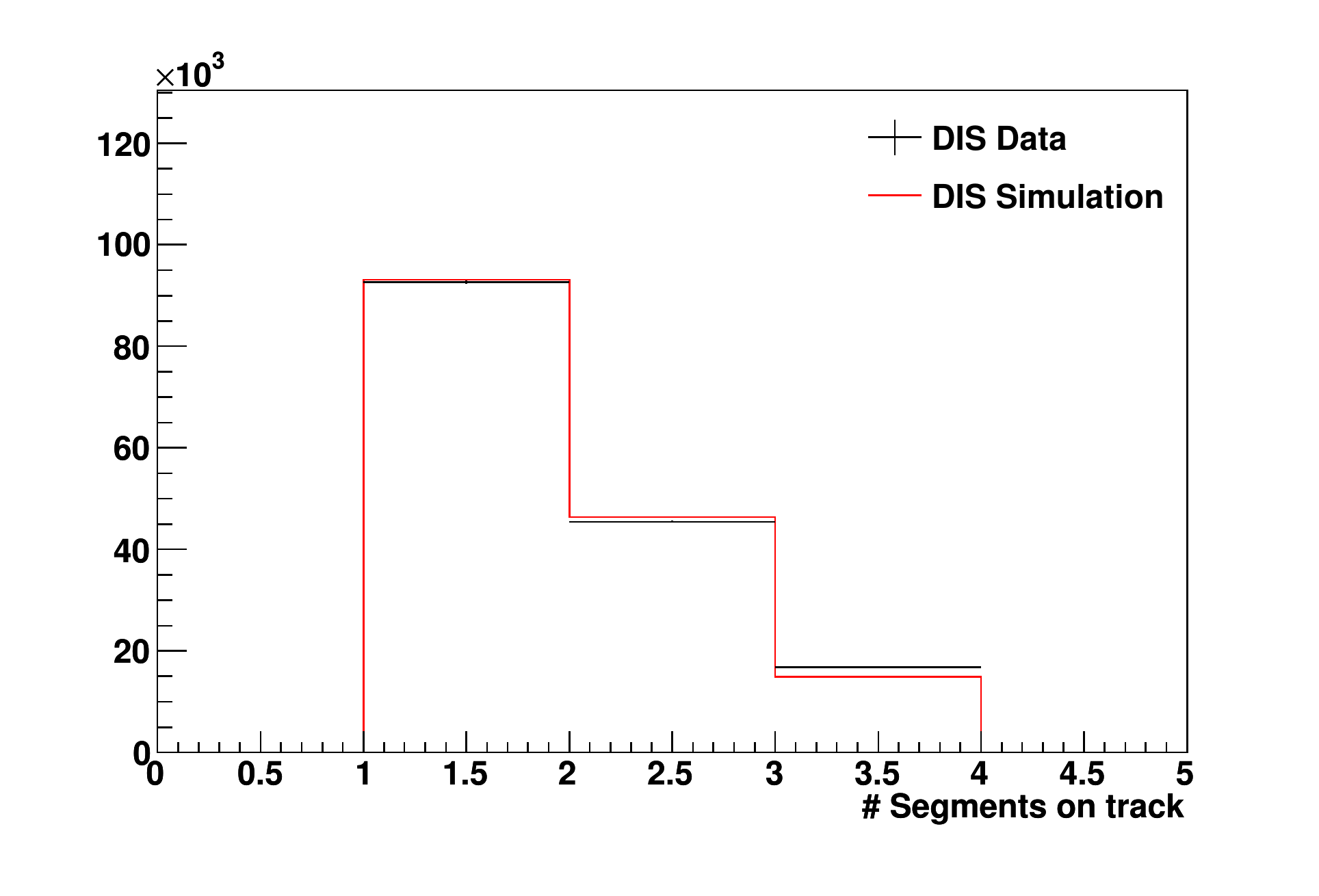}
  \caption{The numbers of clusters and segments on Forward (Combined) tracks, shown
in the left (right) column.}
  \label{fig:incperf2a}
\end{figure}

Forward and Combined tracks which pass the track selection in section
$\ref{sec:tracksel}$ are studied using tracks reconstructed by the
Forward Silicon Tracker\footnote{This is part of the CTD but also
outputs standalone tracks which are used here.} (FST) and CTD
detectors. Track quality selections are applied to tracks from the FST
and CTD and in addition the tracks are required to have $p > 0.5$~GeV.
A Forward (Combined) track is matched to an FST (Central) track if
$|\Delta \theta| < 0.03$~mrad and $|\Delta \phi| < 0.1$~mrad. Figure
$\ref{fig:incperf}$ shows the efficiency of Forward and Combined
tracks as measured by the FST and CTD, respectively. The efficiency is
studied as a function of $\theta$, $\phi$, $p_T$ and track
multiplicity in the FTD.  The efficiency is described to within $10\%$
or better, differentially in every variable, by the simulation.  The
value of the efficiency is low compared to the single-track
reconstruction efficiency of the FTD, which approaches $100\%$ with
the same selection applied (see section $\ref{sec:JPsiStudy}$).  This
can be understood as being a consequence of the high density
environment and is well described by the simulation.  The number of
clusters and segments used in the Forward and Combined tracks is shown
in figure $\ref{fig:incperf2a}$.  It is very well described by the
simulation and the effect of the CTD acceptance can be clearly seen in
the case of Combined tracks.

\begin{figure}
  \centering
  \includegraphics[width=0.49\linewidth]{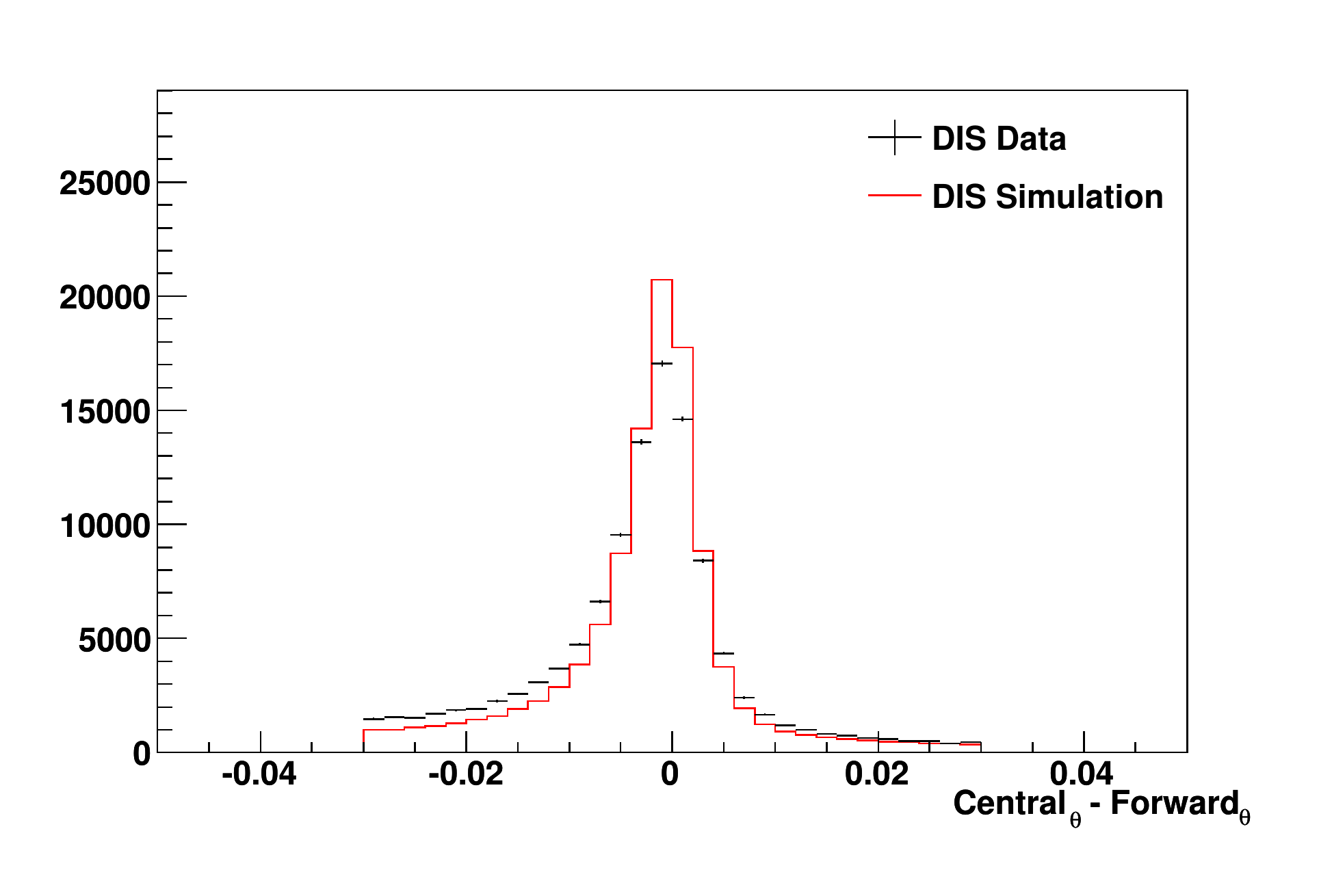}
  \includegraphics[width=0.49\linewidth]{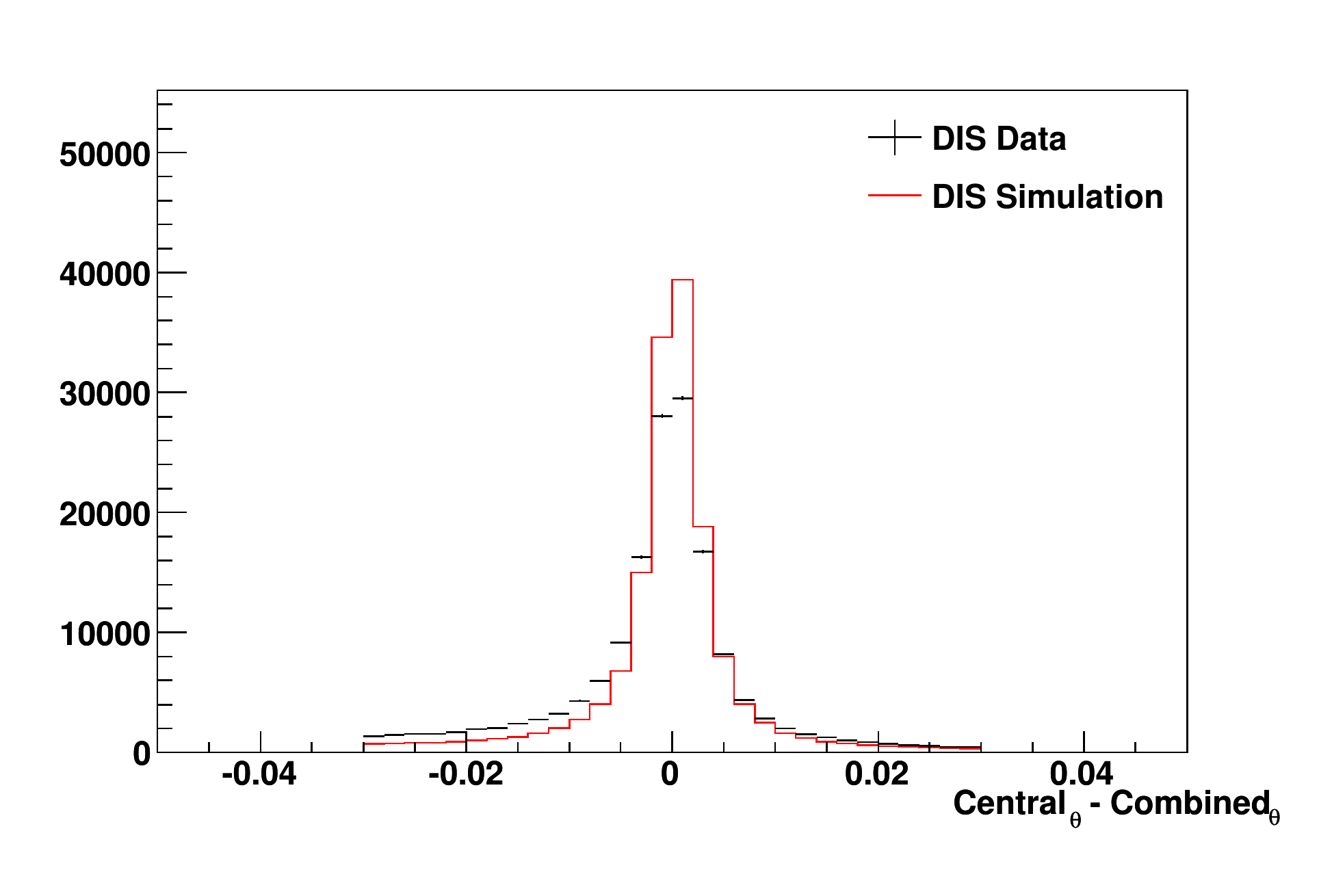}
  \includegraphics[width=0.49\linewidth]{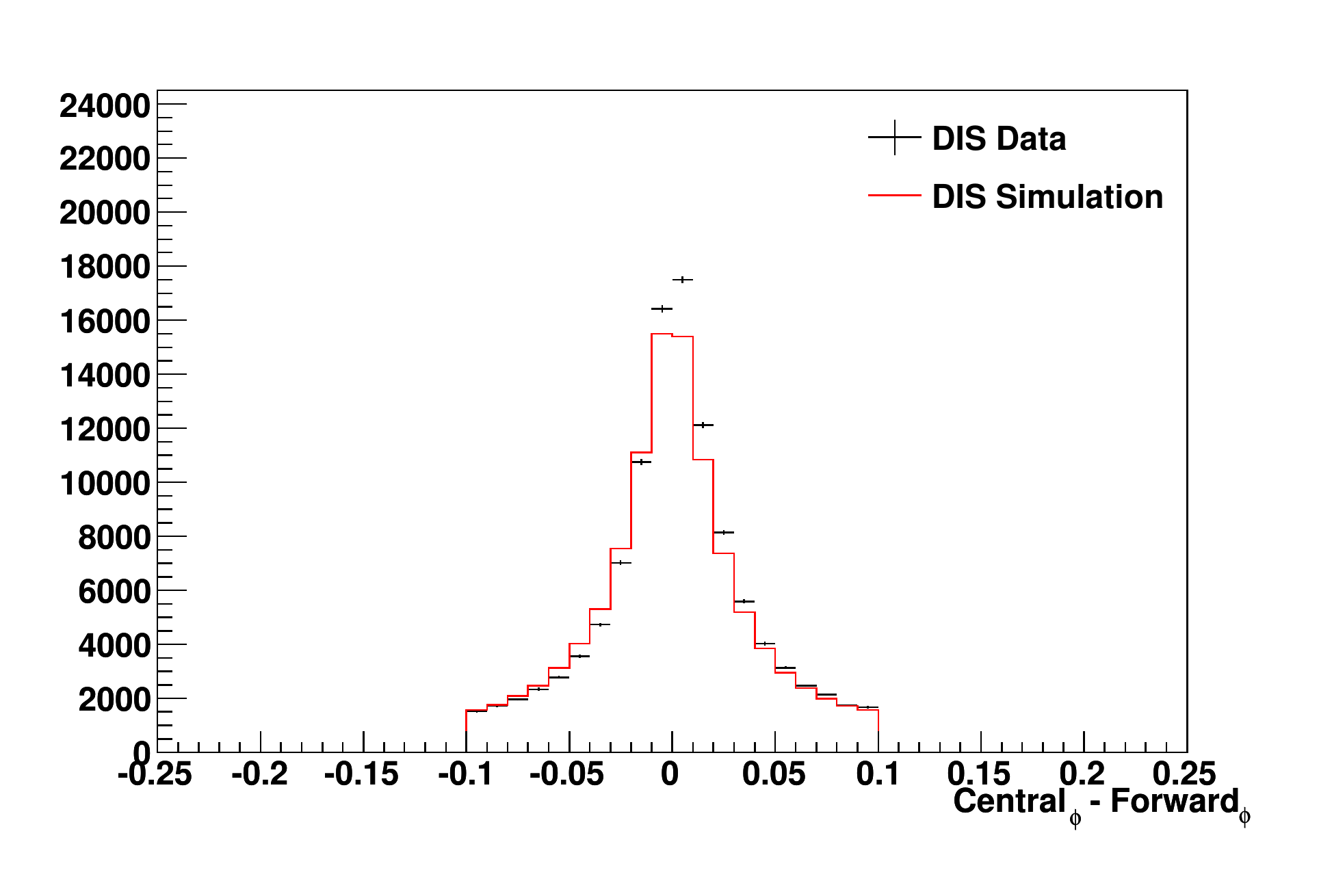}
  \includegraphics[width=0.49\linewidth]{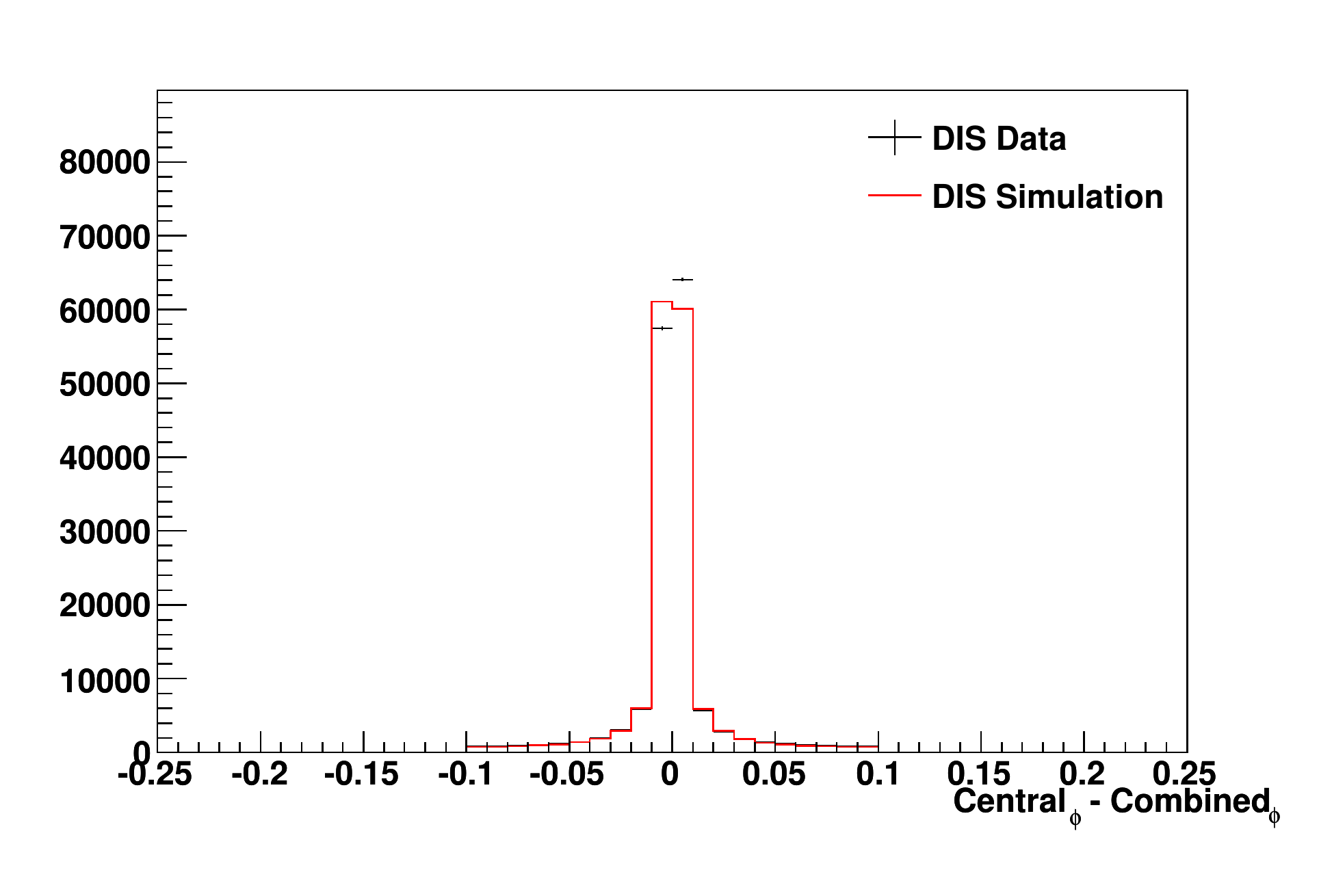}
  \includegraphics[width=0.49\linewidth]{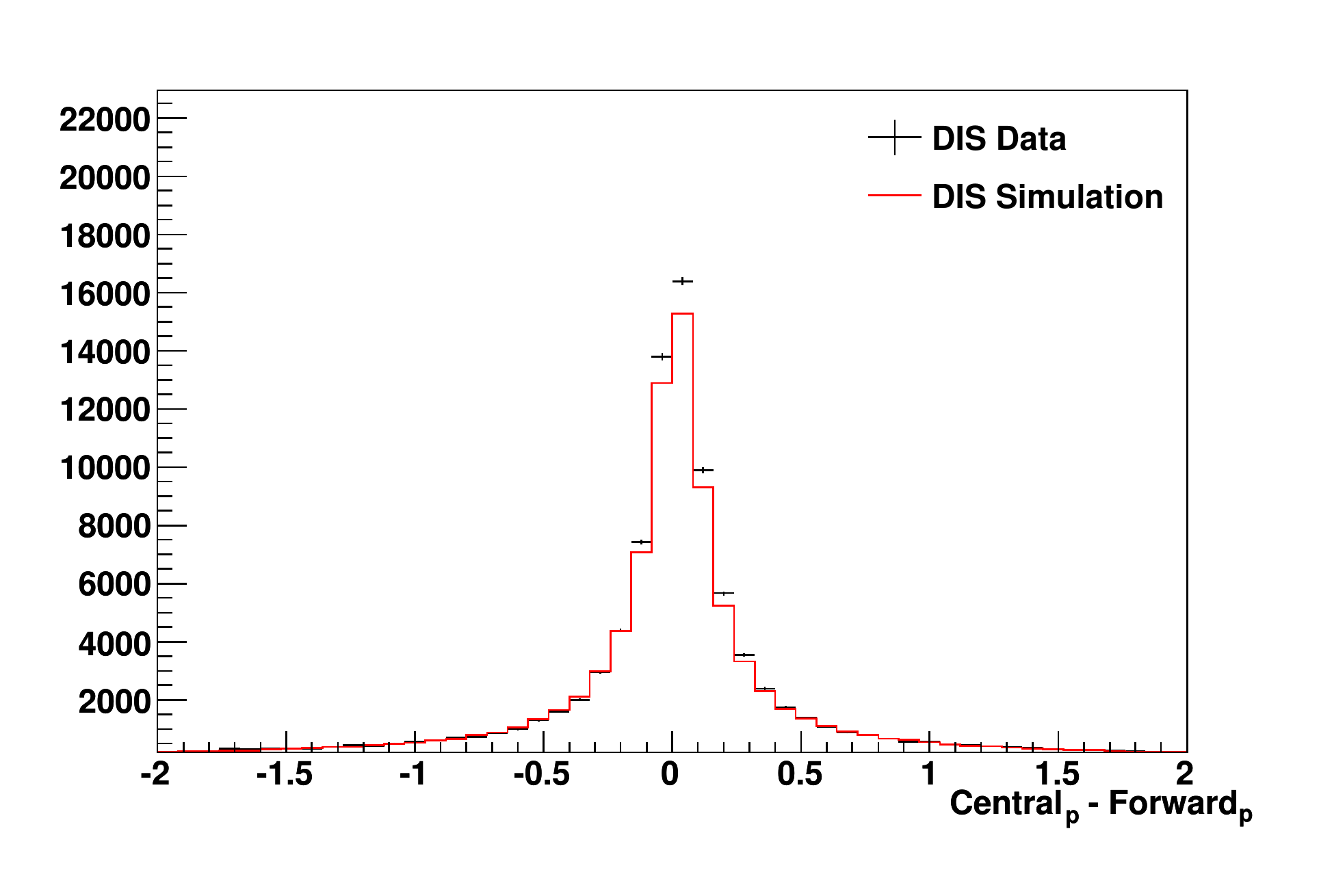}
  \includegraphics[width=0.49\linewidth]{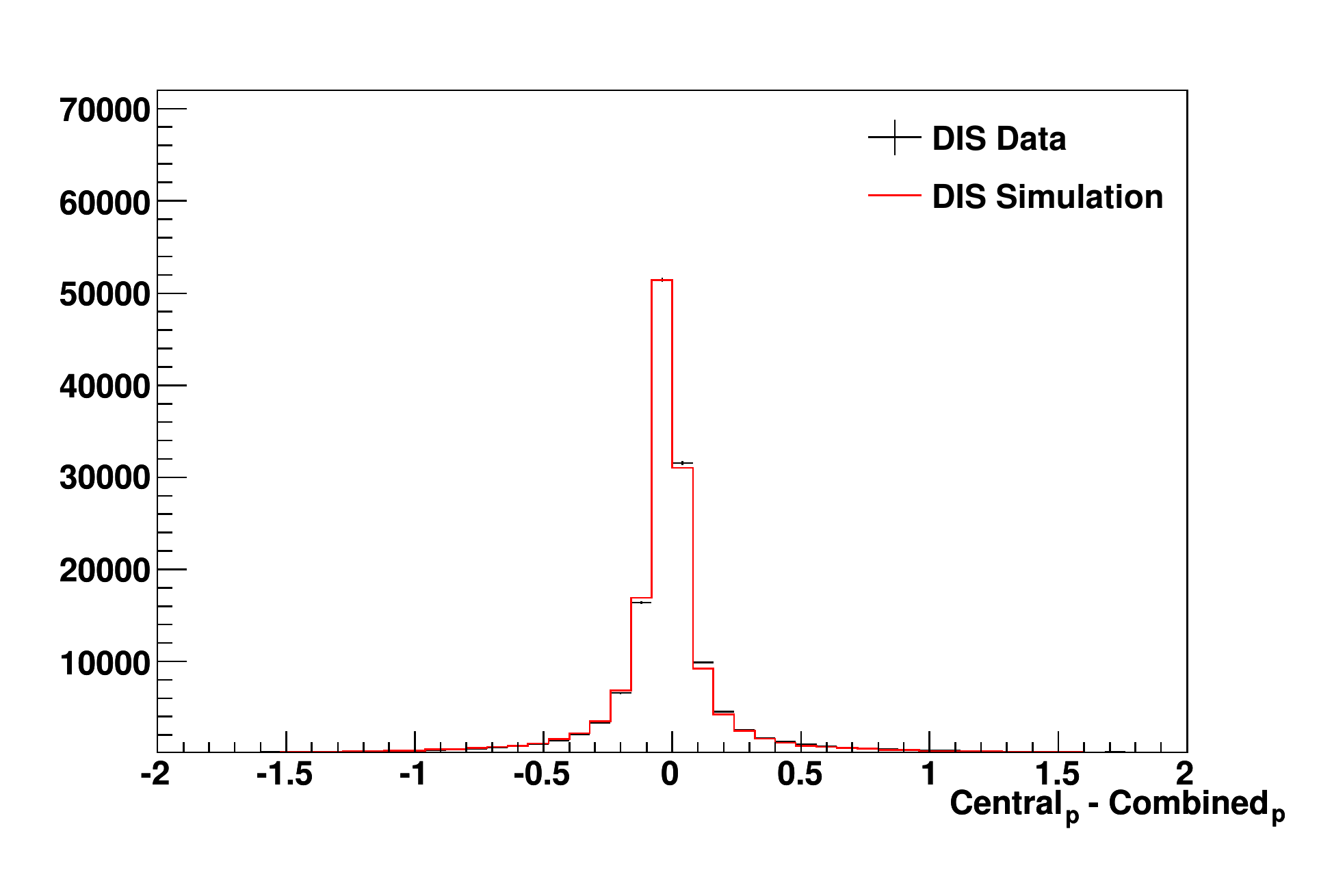}
  \caption{The difference in $\theta$, $\phi$
and $p$ for Forward (Combined) tracks matched to Central tracks in the left (right) column.}
  \label{fig:incperf2}
\end{figure}

The track parameters of Forward and Combined tracks are studied using the CTD, which has a
superior resolution to the FTD. Figure $\ref{fig:incperf2}$ shows the difference in $\theta$, $\phi$
and $p$ for matched tracks. Both $\phi$ and $p$ are in good agreement with
the CTD and this is well described by the simulation, while $\theta$ shows a bias which is slightly larger than
that in the simulation. Studies using the FST show the same
features but the resolution is dominated by the resolution of the
FST. As $\theta$ is the least well constrained parameter in the alignment procedure,
due to the FTD's poor $z$ resolution and the highly non-linear relation
between $z$ residuals and a $\theta$ alignment, this is understood.

\begin{figure}
  \centering
  \includegraphics[width=0.49\linewidth]{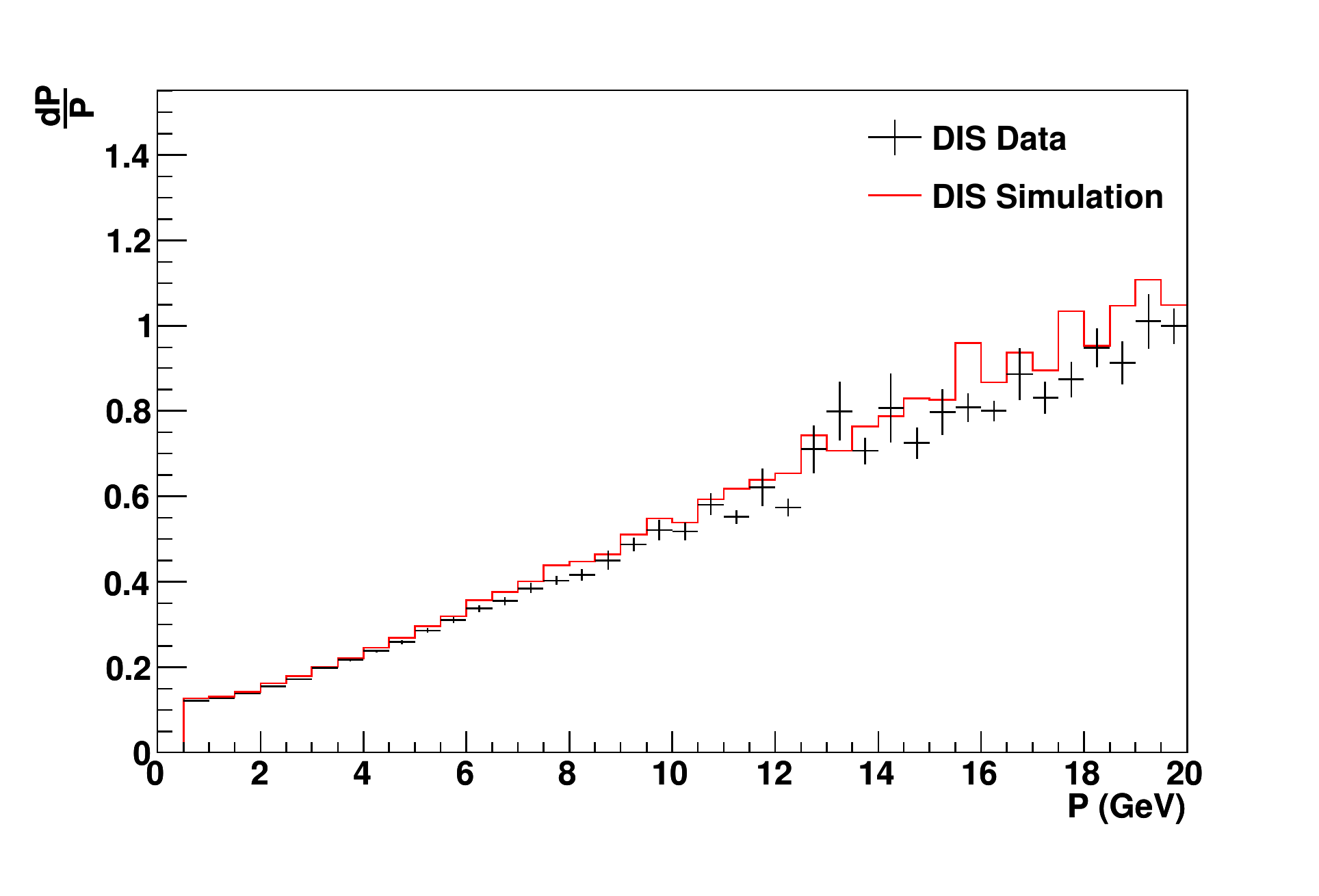}
  \includegraphics[width=0.49\linewidth]{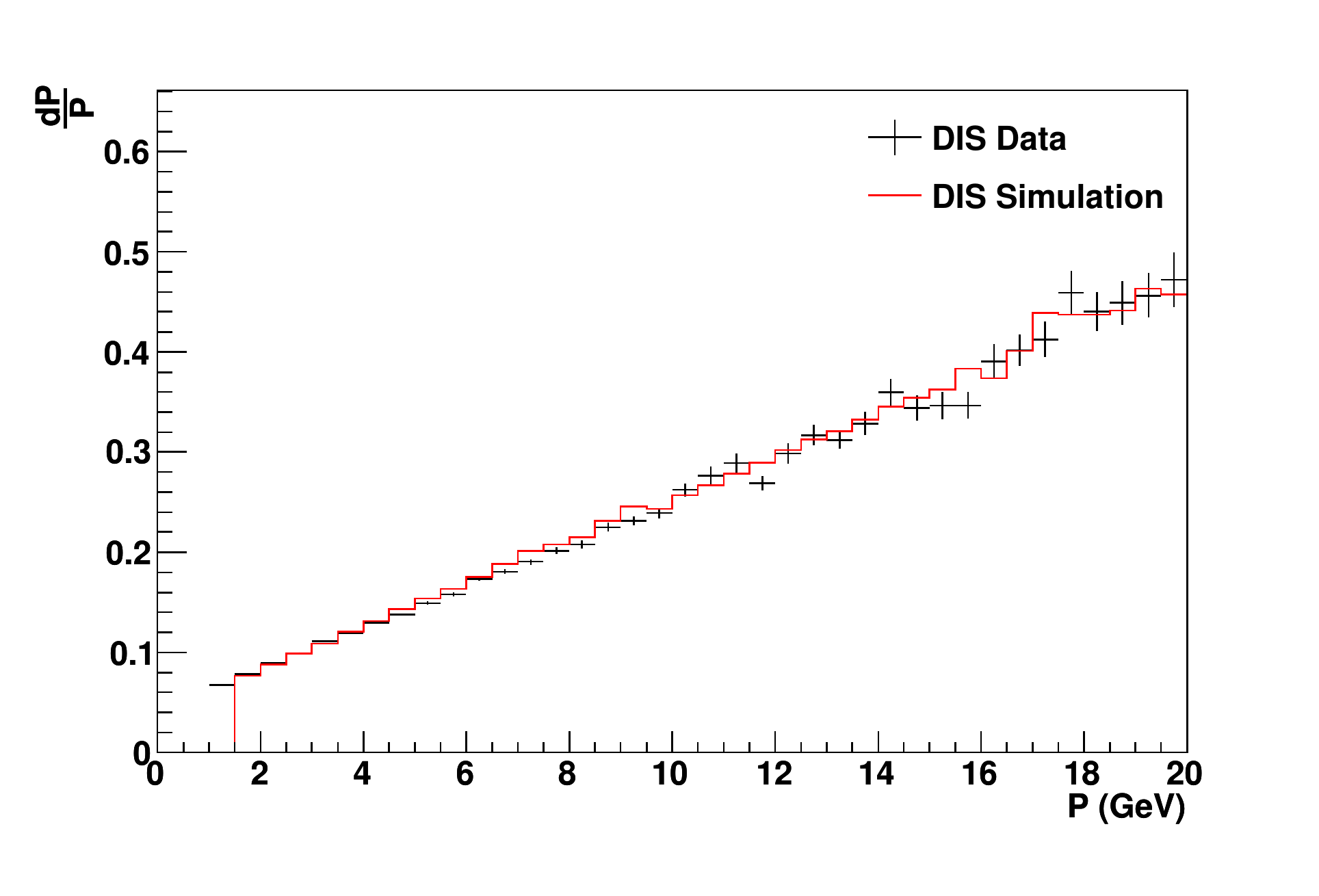}
  \caption{The significance of the momentum measurement $dp/p$ as a function of $p$
  for Forward (left) and Combined tracks.}
  \label{fig:incperf3}
\end{figure}

Finally, the momentum measurement and the error on the momentum
measurement as a function of momentum are shown in figure $\ref{fig:incperf3}$ for both
Forward and Combined tracks. Both track types are described very well by
the simulation with Combined tracks having approximately a factor two
better resolution, as expected. 

\subsection{Comparisons of data and simulation for elastic $J/\Psi$ events}
\label{sec:JPsiStudy}

The hostile environment of the forward
region of H1 in inclusive DIS events, where there is
typically a lot of activity, makes it difficult to cleanly define a
single-track efficiency in data. The elastic $J/\Psi$ process, where the
proton escapes elastically down the forward beam pipe, alleviates this
problem. Furthermore, the di-muon decay channel provides a clean
experimental signature with minimum influence from the abundant dead
material situated in front of the FTD. The Forward Muon
Detector (FMD) and Forward Iron Endcap (FIE) provide independent measurements
of the forward-going muon at low angles, supplementing the CTD.

In the following, $J/\Psi$ candidates are selected with one forward
muon ($\theta < 30^{\circ}$) and one central muon ($\theta >
30^{\circ}$), using the standard H1 muon reconstruction. The elastic
nature of the event is guaranteed by there being no energy deposit in
the H1 calorimeter above $500$ MeV, other than energy associated with
either of the muons. Events are only kept if the two muon tracks are
of opposite charge. The forward muon must have a signal in the FMD or
FIE detectors. The CTD is used to reconstruct the kinematics of the
forward muon in the vast majority of events, but if the CTD fails to
provide a good measurement then these auxiliary detectors are used
instead.  The muon thus reconstructed is referred to as the "Control"
muon.  This allows the efficiency of the FTD to be determined in a
very clean environment but despite this, the track selection given in
section~$\ref{sec:tracksel}$ is applied to Forward and Combined
tracks.  If a selected Forward or Combined track is found, it is used
together with the central muon to reconstruct a $J/\Psi$ candidate,
allowing the momentum scale of the FTD to be studied.

\begin{figure}
  \centering
  \includegraphics[width=0.49\linewidth]{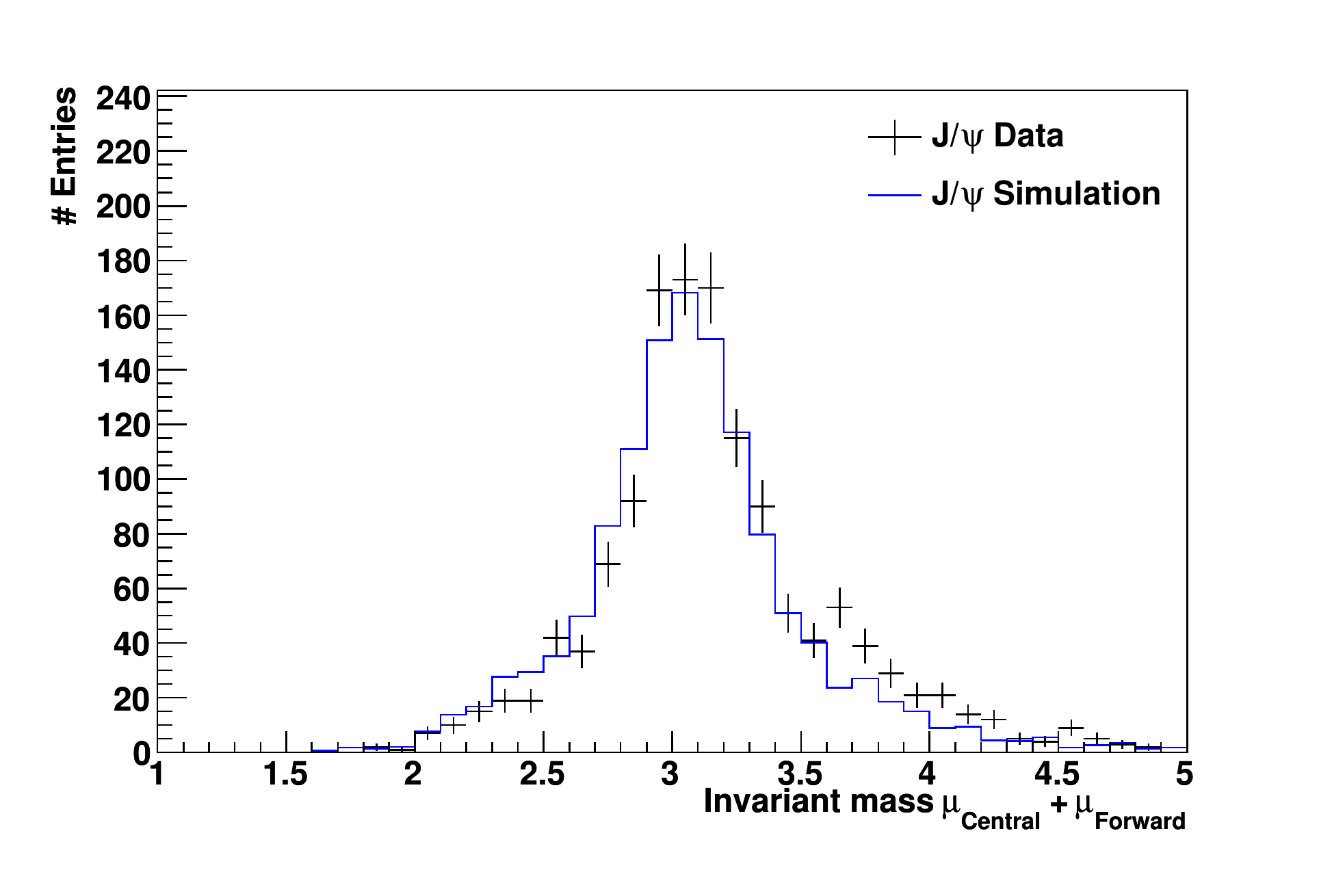}
  \includegraphics[width=0.49\linewidth]{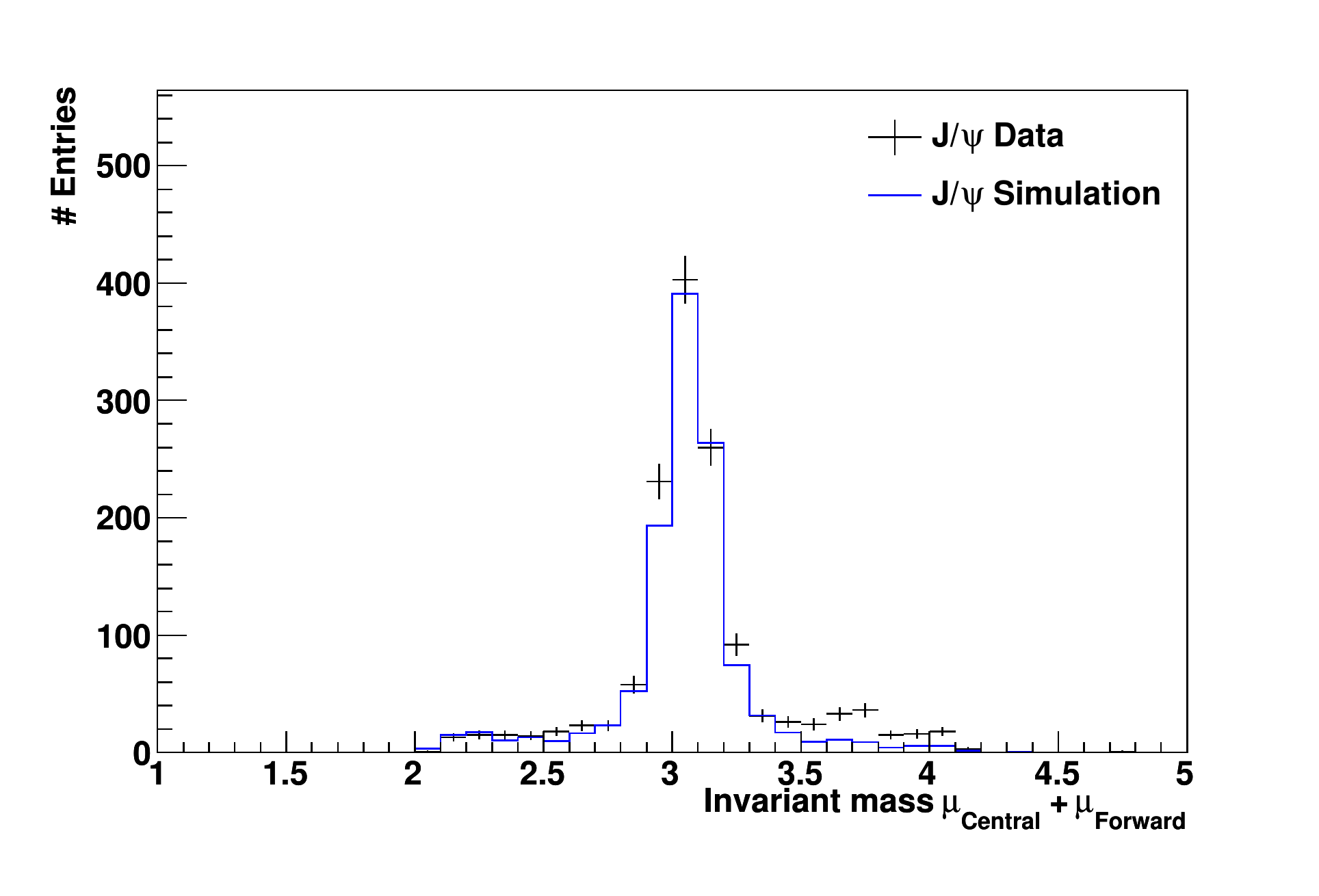}
    \includegraphics[width=0.49\linewidth]{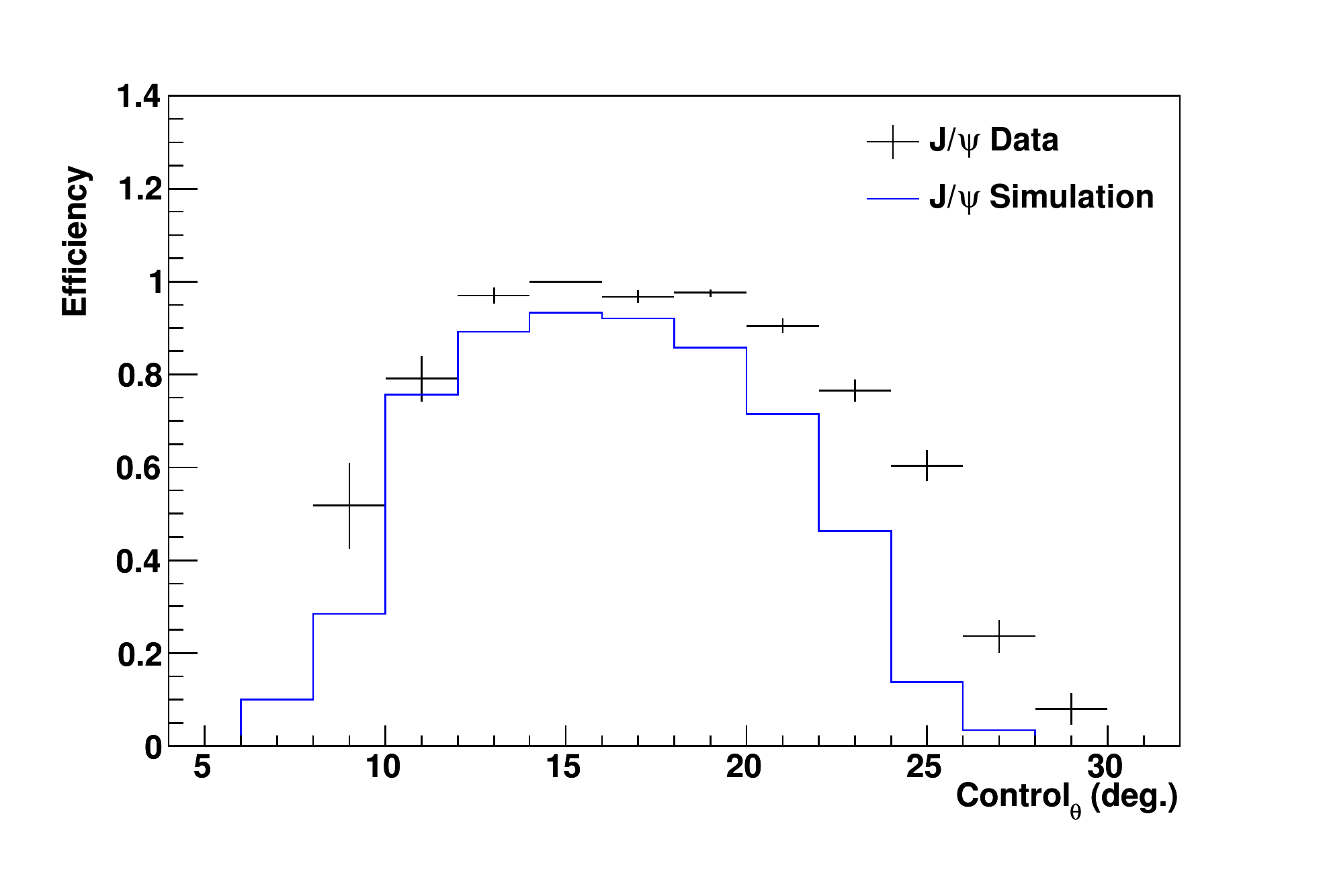}
  \caption{Invariant mass of $J/\Psi$ candidates reconstructed using a Forward (left) or
  Combined (right) track for the forward muon measurement, and (bottom) the efficiency
  of finding either a Forward or a Combined track.}
  \label{fig:jpsi}
\end{figure}

The di-muon mass spectrum using Combined and Forward tracks to
reconstruct the forward muon can be seen in figure $\ref{fig:jpsi}$.
Both data and simulation are in reasonable agreement with the PDG
value for the $J/\Psi$ mass of $3.095$~GeV and a clear $\Psi'$ signal
can also be seen in data. Table $\ref{TAB:JPsi}$ shows the fitted mass
values using a Breit-Wigner function for different track
combinations. The $0.5\%$ underestimate for the control sample
compares reasonably well with the $1\%$ underestimate in the case of
using a Combined or Forward track. The underestimate is well
reproduced by the simulation and is attributed to the lack of detailed
description of the CTD end-wall which is used to calculate the
energy-loss correction. The width of the invariant mass spectrum is
also well described by simulation.

\begin{table}
\begin{center}
\caption{Reconstructed mass of $J/\Psi$ candidates in GeV using a Control, Combined or Forward track
for the forward muon.}
\vspace{0.2cm}
\begin{tabular}{|c|c|c|}
\hline
   & Data & Simulation \\
\hline \hline
Control & $3.08$ & $3.09$ \\ 
Combined & $3.06$ & $3.06$ \\
Forward & $3.07$ & $3.06$ \\
\hline
\end{tabular}
\label{TAB:JPsi}
\end{center}
\end{table}

Figure $\ref{fig:jpsi}$ also shows the efficiency for finding either a
Forward or a Combined track in elastic $J/\Psi$ events.  The
efficiency of the FTD is close to $100\%$ at best in data, while the
simulation tuning applied to describe inclusive DIS data results in a
significant underestimate of the efficiency in the simulation for
these events.  At large $\theta$ the efficiency is shaped by the
detector acceptance, while for $\theta < 10^{\circ}$ the efficiency
downturn is entirely the result of the track selection.  If the tuning
is not applied to the simulation, the efficiency is very well
described.

The importance of this elastic $J/\Psi$ sample in understanding the
performance of the FTD cannot be over-stated.  The clean environment
and clear physics signal, together with using other detectors to
unambiguously and independently find the signal in data, allowed for
great improvements and testing of the FTD algorithms.  In particular,
the original tolerances used to form segment candidates, described in
section $\ref{subsubsec:segmentform}$, were improved using this
sample.  This resulted in a $15\%$ higher track-finding efficiency and
$25\%$ higher efficiency for finding tracks with more than one segment,
compared to the original values taken from Monte Carlo studies.

\section{Summary}

The upgraded Forward Track Detector of H1 for the HERA II phase has
been presented.  The detector hardware was significantly modified in
order to improve the detector performance.  Upon commissioning the
detector, ageing was found to have affected the old Planar chambers
that were still in use.  A new QT algorithm was developed to allow the
chambers to be operated at lower gain and mitigate the effects of
ageing, which successfully restored the single hit efficiency to
$\sim96\%$, allowing the chambers to be operated at close to design
parameters for the rest of HERA running, with an uptime of $\sim98\%$.

The software of the FTD was rewritten to take advantage of the
improved hardware.  The performance of the new algorithms using the
new hardware configuration surpassed the previous incarnation of the
FTD, with very high cluster and segment finding efficiencies achieved
together with high purity.  The Kalman filter used to produce the
final tracks from the reconstruction was modified to further improve
the track reconstruction, adding and rejecting clusters to improve the
overall track quality of Forward tracks.  These Forward tracks were
combined with central tracking information to provide Combined tracks
with improved precision, while providing a framework to align the FTD
to the rest of H1.

The performance of the FTD was measured using several independent
detectors, chiefly the CTD.  In inclusive NC DIS events, the
description of Forward and Combined tracks, after a track selection,
by the H1 simulation (tuned to match the observed cluster-finding
efficiency in data) is good.  While the efficiency of the FTD for
reconstructing Forward and Combined tracks is relatively low ($\sim
30-50\%$) in this high multiplicity environment, it is well described
by the simulation, also differentially.  The track parameters of
selected Forward and Combined tracks are accurate and are again well
described by simulation, allowing these tracks to be used in the H1
HFS algorithm.  Finally, using a sample of elastic $J/\Psi$ events,
the FTD was shown to have an accurate and well simulated momentum
measurement and a single-track efficiency of close to $100\%$.

\end{document}